\documentclass[12pt]{article}
\pdfoutput=1
\usepackage{graphicx}
\newcommand{\nn}{\nonumber}
\newcommand{\nc}{\newcommand}  
\nc{\del}{\partial}
\nc{\beq}{\begin{equation}}  
\nc{\eeq}{\end{equation}}  
\nc{\be}{\begin{equation}}
\nc{\ee}{\end{equation}}
\nc{\beqa}{\begin{eqnarray}}  
\nc{\eeqa}{\end{eqnarray}}  
\nc{\bea}{\begin{eqnarray}}  
\nc{\eea}{\end{eqnarray}}  
\nc{\ra}{\rightarrow}  
\nc{\lsim}{\begin{array}{c}\,\sim\vspace{-21pt}\\< \end{array}}  
\nc{\gsim}{\begin{array}{c}\sim\vspace{-21pt}\\> \end{array}}  
\nc{\half}{{1 \over 2}}

\nc{\Kahler}{\textrm{K\"{a}hler}~}
\nc{\LL}{L}  
\nc{\vv}{\tilde{v}}  
\nc{\GG}{\widetilde{G}}  
\nc{\MM}{\ensuremath{\mathcal{M}}}  
\nc{\UU}{\ensuremath{\mathcal{U}}}  
\nc{\ZZ}{\ensuremath{\mathcal{{\cal{Z}}}}}  
\nc{\Mu}{\ensuremath{M_{u}}}
\nc{\Md}{\ensuremath{M_{d}}}
\nc{\Mtu}{\ensuremath{\tilde{M}_{u}}}
\nc{\Mtd}{\ensuremath{\tilde{M}_{d}}}

\title{  
\vspace*{-2.3cm}  
\begin{flushright}  
\end{flushright}  
\vspace{1.5cm}  
\Large  
\textbf{The Supersymmetric Higgs}\vspace*{1.0cm}   
\author{\large  
\textbf{Puneet Batra},  
and  
\textbf{Eduardo Pont\'{o}n}
\\
\normalsize\emph{Department of Physics, Columbia University,}\\
\normalsize\emph{538 W. 120th St, New York, NY 10027, USA} \\  
}
\date{}}  
\begin{document}  
\setcounter{page}{0}  
\maketitle  
%\date{}  
%\vspace*{1cm}  
\begin{abstract}   
In the Minimal Supersymmetric Standard Model (the MSSM), the
electroweak symmetry is restored as supersymmetry-breaking terms are
turned off.  We describe a generic extension of the MSSM where the
electroweak symmetry is broken in the supersymmetric limit.  We call
this limit the ``sEWSB'' phase, short for supersymmetric electroweak
symmetry breaking.  We define this phase in an effective field theory
that only contains the MSSM degrees of freedom.  The sEWSB vacua
naturally have an inverted scalar spectrum, where the {\it heaviest}
CP-even Higgs state has Standard Model-like couplings to the massive
vector bosons; experimental constraints in the scalar Higgs sector are
more easily satisfied than in the MSSM.
\end{abstract}  
\thispagestyle{empty}  
\newpage  
  
\setcounter{page}{1}

\baselineskip18pt

%-----------------------------------------------------------------------------
\section{Introduction}
\label{sec:intro}  
%-----------------------------------------------------------------------------
%
The Minimal Supersymmetric extension of the Standard Model (MSSM)
provides a framework for understanding the origin of electroweak
symmetry breaking (EWSB).  The Higgs fields will acquire vacuum
expectation values (VEV's) only if their mass parameters live in a
window that produces a non-trivial but stable global minimum in the
Higgs potential.  This window always requires supersymmetry
(SUSY)-breaking and may occur radiatively \cite{Ibanez:1982fr}.

Of the two neutral CP-even states in the MSSM, typically the lightest
CP-even state couples to the massive W and Z vector bosons like the
Standard Model Higgs (is ``SM-like'').  At tree-level, this state has
a mass lighter than $m_Z$ because the Higgs potential is stabilized by
\Kahler terms proportional to the electroweak (EW) gauge couplings.  As is
well known, large SUSY-breaking effects in the stop-top sector can
allow this SM-like Higgs state to escape LEP-II bounds, but only at
the cost of tuning the parameters of the theory.

However, if EWSB occurs instead in the supersymmetric limit, it is the
non-SM-like Higgs CP-even state whose mass is tied to $m_Z$, {\it not}
the SM-like Higgs.  The SM-like Higgs state is part of a chiral
supermultiplet whose mass is not related to the electroweak gauge
couplings and is not related to $m_Z$ at tree-level.  We call any
vacuum in which the electroweak symmetry remains broken as
SUSY-breaking is turned off a ``supersymmetric electroweak symmetry
breaking'' vacuum (sEWSB vacuum).  Considering again the LEP-II
bounds, the most interesting feature of sEWSB vacua is that the
CP-even scalar spectrum may be inverted compared to the usual spectrum
of the MSSM: the {\it heavier} CP-even state, not the lighter, is the
SM-like Higgs field.  In the MSSM, it is possible to have viable inverted
CP-even spectra but only with large radiative corrections.

Further, sEWSB will occur with only the mild assumption of a new
approximately supersymmetric physics threshold just above the
weak-scale that couples to the MSSM Higgs fields.  We can therefore
understand sEWSB most simply by working in an effective theory that
only contains the MSSM degrees of freedom and additional
non-renormalizable interactions.  Focusing on the Higgs sector of the
theory, the most general superpotential that can arise from
integrating out a \textit{supersymmetric} threshold at the scale
$\mu_{S}$ is
\beqa
W &=& \mu H_{u} H_{d} + \frac{\omega_{1}}{2\mu_{S}} (H_{u} H_{d})^{2} + 
\frac{\omega_{2}}{3\mu^{3}_{S}} (H_{u} H_{d})^{3} + \cdots~,
\label{non-renorm-W}
\eeqa
where we have suppressed the $SU(2)_L$ indices and $H_u H_d = H_u^+
H_d^- - H_u^0 H_d^0$.  The ellipses represent terms suppressed by
higher powers of the scale $\mu_{S}$, and the $\omega_{i}$ are
dimensionless coefficients.  Keeping only the first two terms, for
simplicity, we see that the $F$-flatness conditions are
satisfied by the origin in field space, and also by a nontrivial VEV,
\bea
\langle H_u H_d \rangle =  -\mu \mu_S / \omega_1~.
\label{eq:susymin}
\eea 
Thus, the EW scale may arise as the geometric mean of the $\mu$-term
and the scale of some relatively heavy new physics and have a purely
supersymmetric origin.  As we show in
Subsection~\ref{sec:SUSYSpectrum}, the spectrum of this vacuum is very
simple: most of the Higgs fields (scalar and fermion components) are
``eaten'' by the vector superfields and together have masses equal to
$m_W$ or $m_Z$.  One neutral Higgs superfield remains, which contains
the SM-like Higgs, with mass $2 |\mu|$.  For $2 |\mu| > m_Z$, the
scalar spectrum is inverted compared to the decoupling limit of the
MSSM: the light CP-even state with mass $m_Z$ is not SM-like, while
the heavy CP-even state is at $2 |\mu|$ and is SM-like.

Since we are working in a non-renormalizable theory, it is not enough
that sEWSB occurs, we require that the effective field theory (EFT)
remain valid in an expansion around this minimum---all ignored
operators beyond the first two should give only small corrections to
our analysis.  Supersymmetry plays a prominent role in maintaining the
validity of the EFT. Non-renormalizable operators either in the
\Kahler potential or in the superpotential are suppressed by
\bea 
{\langle H \rangle^2 \over \mu_S^2} \sim {2
\over \omega_1} {\mu \over \mu_S}~, 
\label{EFTvalidity}
\eea 
and can be self-consistently ignored provided $\mu \ll \mu_S$.  SUSY
drives this suppression in two ways.  First, the separation of scales
between $\mu$ and $\mu_S$ is technically natural in a supersymmetric
theory.  Second, the sEWSB VEV results from balancing a dimension-6
term in the scalar potential against a dimension-4 term, so that
$\langle H^2 \rangle$ is proportional to the Higgs' quartic times the
non-renormalizable scale, $\mu_S^2/\omega^2$.  However, holomorphicity
and gauge invariance in the superpotential only allows a quartic term
of order $\omega \mu/\mu_S$ along the Higgs D-flat direction in the
scalar potential.  If any larger quartic were allowed, the validity of
the EFT would be ruined.  Ironically, the absence of a large quartic
in the Higgs superpotential is exactly why there is a little hierarchy
problem in the MSSM to begin with.

Given the bounds from direct searches on superpartners, SUSY must be
broken, and we expect the SUSY limit to be deformed by soft-masses of
order the electroweak scale.  We incorporate the effects of
SUSY-breaking in Section~\ref{sec:SUSYBreaking} and show how to
consistently identify sEWSB vacua in this limit.  Depending on the
parameter choice, the sEWSB minimum of Eq.~(\ref{eq:susymin}) may be
the only non-trivial minimum of the theory, or it can be joined by a
vacuum which is continuously connected to the usual EWSB vacuum of the
MSSM in the limit that the non-renormalizable operators of
Eq.~(\ref{non-renorm-W}) are turned off (MSSM-like vacua).  Even with
SUSY-breaking turned on, we show in Section \ref{sec:pheno} that sEWSB
vacua can share the qualitative features of the pure SUSY-limit: the
heavier CP-even state has SM-like Higgs couplings to massive vector
bosons.

One of the main phenomenological tensions in this vacuum is the forced
separation between $\mu$ and $\mu_S$.  This ratio should be small, to
maintain control of the effective theory, but there is a tension
between making $\mu_S$ large while keeping the ratio $\mu \mu_S \sim
v^2$ fixed.  The SUSY-limit forces the charginos to have mass $m_W$.
Pushing these states above LEP-II bounds requires keeping $\mu$ as
large as possible when SUSY is broken.  In Section \ref{sec:pheno}, we
show that charginos and neutralinos near the LEP-II bound are a fairly
generic prediction of sEWSB vacua, and that the lightest chargino may
be lighter than the lightest neutralino (the gravitino could be the
LSP in this case).  This NLSP chargino would lead to an enhanced set
of $W$ bosons in cascade decays \cite{Kribs:2008hq}.

In Section~\ref{sec:UV} we discuss one of the simplest ultraviolet
completions that can lead to sEWSB vacua: adding a singlet superfield
$S$ to the MSSM, with a supersymmetric mass $\mu_S$ and a trilinear $S
H_u H_d$ coupling.  Unlike the NMSSM \cite{Nilles:1982dy}, we do not
explain the origin of the $\mu$--term in the MSSM: this UV theory
includes an explicit $\mu H_u H_d$ term.  It is well-known that the
LEP-II limit can also be escaped by integrating out a singlet
superfield in the {\it non}-SUSY limit \cite{Espinosa:1992hp}; here we
assume $\mu_S$ is much larger than the scale of SUSY-breaking.  The
Fat Higgs model~\cite{Harnik:2003rs} is another example of a
singlet-extended MSSM theory that exhibits EWSB in the SUSY-limit, but
is not described by our EFT, since the field $S$ cannot be decoupled
from the spectrum in a supersymmetric limit.  The singlet UV
completion of our theory belongs to the more general analyses of
theories with singlet superfields and the coupling $\lambda S H_u H_d$
\cite{Nomura:2005rk}.
 
An EFT approach to parameterize extensions to the MSSM up to terms of
${\cal O}(H^4)$ in the superpotential has already been used to analyze
the effects of the leading, renormalizable, ${\cal O}(H^4)$ terms in
the scalar potential \cite{Brignole:2003cm}.  These analyses are
useful for calculating perturbations to MSSM-like vacua.  The sEWSB
vacua require keeping terms of order ${\cal O}(H^4)$ in the
superpotential {\it and the full set} of ${\cal O}(H^6)$ terms in the
scalar potential that are generated by the superpotential, a case not
seriously considered in previous studies.

%-----------------------------------------------------------------------------
\section{Supersymmetric Electroweak Symmetry Breaking}
\label{sec:SUSYEWSB}  
%-----------------------------------------------------------------------------

 As we will see, the qualitative physical properties of the sEWSB
 vacuum can already be understood in the supersymmetric limit.  It is
 therefore useful to study in some detail the physics of EWSB when
 SUSY is exact, which we do in this section.  We consider the effects
 of SUSY breaking, under the assumption that the heavy threshold
 $\mu_{S}$ is approximately supersymmetric, in
 Section~\ref{sec:SUSYBreaking}.

%-----------------------------------------------------------------------------
\subsection{Validity of the Effective Theory on the sEWSB Vacuum}
\label{sec:EFT}  
%-----------------------------------------------------------------------------

Our main observation is that in the presence of the higher-dimension
operators in the superpotential of Eq.~(\ref{non-renorm-W}) there is a
non-trivial ground state that can be reliably studied within the EFT
framework.  The only condition is that there exists a mild hierarchy
between $\mu$ and the new physics threshold $\mu_{S}$.

Indeed, assuming that the first non-renormalizable operator in
Eq.~(\ref{non-renorm-W}) is non-vanishing, the $F$-flatness conditions
can be satisfied both at the origin of field space and at a VEV of
order $\mu\mu_{S}/\omega_{1}$.  This solution exists for any sign of
the dimensionless coefficient $\omega_{1}$.  It is a solution to the
$F$-flatness conditions where the two leading terms in
Eq.~(\ref{non-renorm-W}) approximately cancel, while the remaining
operators give contributions that are suppressed by powers of
$\mu/\mu_{S}$ (times ratios of dimensionless coefficients).  Thus, we
can capture the physical properties of this vacuum to leading order in
$\mu/\mu_{S}$ by keeping the first two terms in
Eq.~(\ref{non-renorm-W}).  This defines the zeroth order
approximation.  Operators in the superpotential suppressed by
$1/\mu^{2n+1}_{S}$ with $n \geq 1$, give corrections to physical
observables that are suppressed by at least $(\mu/\mu_{S})^{n}$, which
we refer to as the n-th order approximation.  Notice that the
importance of an operator, whether non-renormalizable or not, depends
on the vacuum state one is expanding field fluctuations about.  In
general, to estimate the relevance of any operator one should do the
power counting after expanding around the VEV of interest.

One might also worry about the effects of higher-dimension operators
in the \Kahler potential.  However, these enter at
\textit{next-to-leading} order in the $1/\mu_{S}$ expansion, e.g.
\beqa
K = H^{\dagger}_{u} \, e^{V} H_{u} \left[ 1 + \frac{1}{\mu^2_{S}} f_{u} \right] + 
H^{\dagger}_{d} \, e^{V} H_{d} \left[ 1 + \frac{1}{\mu^2_{S}} f_{d} \right] + 
\frac{c_{1}}{\mu^{2}_{S}} \, |H_{u} \epsilon H_{d}|^{2} + \cdots~,
\label{Kahler}
\eeqa
where 
\beqa
f_{u} &=& \frac{1}{2} a^{u}_{1} \, H^{\dagger}_{u} \, e^{V} H_{u} + 
\frac{1}{2} a^{ud}_{1} \, H^{\dagger}_{d} \, e^{V} H_{d} + 
\left( b^{u}_{1} \, H_{u} H_{d} + {\rm h.c.} \right) + {\cal O}\left( \frac{1}{\mu^2_{S}} \right)~,
\label{fu} \\ [0.6em]
f_{d} &=& \frac{1}{2} a^{d}_{1} \, H^{\dagger}_{d} \, e^{V} H_{d} + 
\frac{1}{2} a^{ud}_{1} \, H^{\dagger}_{u} \, e^{V} H_{u} + 
\left( b^{d}_{1} \, H_{u}  H_{d} + {\rm h.c.} \right) + {\cal O}\left( \frac{1}{\mu^2_{S}} \right)~.
\label{fd}
\eeqa
Their effects on the physical properties of the vacuum of
Eq.~(\ref{eq:susymin}) are also suppressed by $\mu/\mu_{S}$ and
correspond to small corrections to the zeroth order solution described
in the previous paragraph.\footnote{\Kahler terms suppressed by
$1/\mu^{2n}_{S}$ give corrections suppressed by at least
$(\mu/\mu_{S})^{n}$.} For instance, although the leading order
$D$-terms imply that $\tan\beta = \langle H_{u} \rangle/\langle H_{d}
\rangle = \pm 1$, the higher-dimension \Kahler corrections can lead to
$|\tan\beta| \neq 1$ if $a^{u}_{1} \neq a^{d}_{1}$, or $b^{u}_{1} \neq
b^{d}_{1}$, etc.  [see Eqs.~(\ref{VD}), (\ref{Ds}) and
(\ref{NonminimalDs}) in Appendix~\ref{app:kahlermetric} for the
general expressions of the $D$-term potential].  However, to the
extent that $\mu/\mu_{S}$ is small, one finds that $|\tan\beta|$
remains close to one in the SUSY limit.  Nevertheless, the \Kahler
terms can have other phenomenologically relevant effects that are
pointed out in Subsection~\ref{sec:mixing}.  There may also be terms
containing SUSY covariant derivatives that we do not show explicitly,
since they lead to derivative interactions that do not affect the
vacuum or spectrum of the theory.

\begin{figure}[t]  
\centerline{\includegraphics[width=4 in]{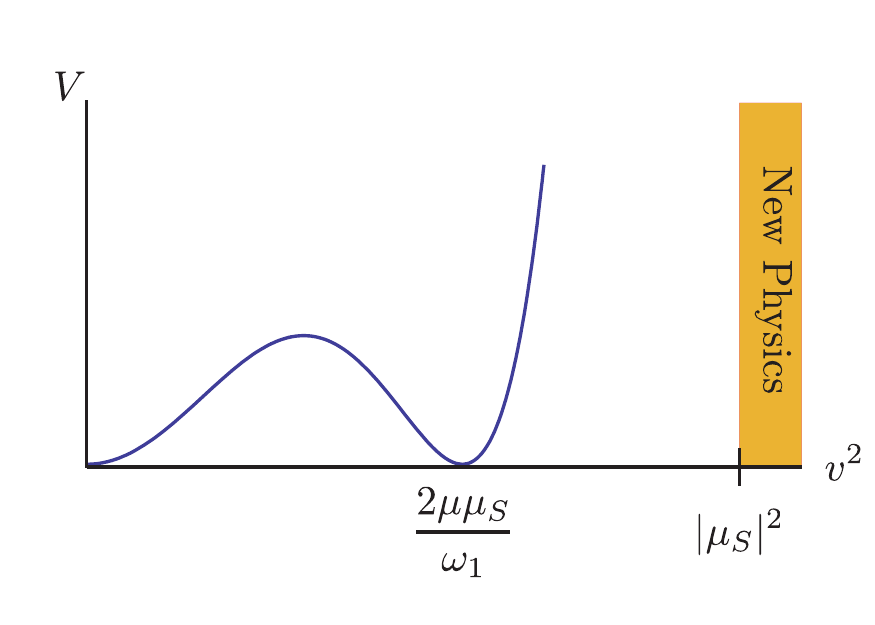}}
\caption{The phase structure of the superpotential in
Eq.~(\ref{non-renorm-W}) keeping only the leading correction, along
the $\tan\beta=1$ slice.  Supersymmetry allows us to reliably
calculate around the EWSB minima, since the scale of new physics may
be much larger than all other mass scales in the effective theory.}
\label{fig:susy_phase}  
\end{figure} 

In summary, it is possible to study the properties of the sEWSB vacuum
from Eq.~(\ref{non-renorm-W}) without a complete specification of the
physics that gives rise to the tower of higher-dimension operators, so
that an EFT analysis is appropriate.  In particular, the theory that
includes the higher-dimension operators has at least two degenerate
SUSY-preserving minima: the origin and a vacuum where EWSB occurs.
These supersymmetric vacua are degenerate and separated by a potential
barrier as shown schematically in Fig.~\ref{fig:susy_phase}.  We can
characterize the sEWSB minimum by
\bea
\langle H_u^0 \rangle \approx \langle H_d^0 \rangle \approx  \sqrt{\mu \mu_S / \omega_1}~,
\label{eq:susymin2}
\eea
which holds up to corrections of order $\mu/\mu_{S}$.  Here we have
used a combination of $SU(2)_L \times U(1)_Y$ gauge transformations to
make both VEV's real, positive, and in the electrically neutral
components, together with an additional field redefinition to make the
quantity $\mu \mu_S/ \omega_1$ real and positive.  In the following we
will refer to the vacuum of Eq.~(\ref{eq:susymin2}) as the ``sEWSB
vacuum'' (for supersymmetric EWSB vacuum).

One might still wonder if other non-trivial vacua exist when the
superpotential has the form of Eq.~(\ref{non-renorm-W}).  In general,
except for the sEWSB vacuum described above, all other potential
solutions to the $F$-flatness conditions would correspond to VEV's of
order $\mu_{S}$, and are therefore outside the realm of the EFT. In
fact, the question of whether such vacua actually exist or not can
only be answered within the context of a given UV completion.  It is
logically possible that additional solutions with VEV's parametrically
smaller than $\mu_{S}$ exist, but this can only happen for special
choices of the coefficients $\omega_{i}$.  For example, solutions that
arise from balancing the $\mu$-term with an $\omega_{n}$-operator [the
operator with coefficient $\omega_{n}$ in Eq.~(\ref{non-renorm-W})]
exist only if the coefficients of all $\omega_{i}$-operators with $i <
n$ are suppressed by appropriate powers of $\mu/\mu_{S}$.  This latter
quantity has to be small in order that the $\omega_{i}$-operators with
$i > n$ can be neglected.  In particular, if the $\omega_{1}$-operator
is generated by the physics at $\mu_{S}$ with a coefficient larger
than ${\cal O}(\mu/\mu_{S})^{1/2}$, no such solutions exist.  We also
assume here that the $\omega_{n}$ are smaller than the NDA estimate
$(16\pi^2)^n$ \cite{Luty:1997fk}.  If the physics at $\mu_{S}$ is
strongly coupled, our analysis cannot reliably establish the existence
of non-trivial minima in the SUSY limit.  However, notice that due to
non-renormalization theorems, it is possible that all but a finite
number of operators in the superpotential vanish.

In this paper, we concentrate on the sEWSB vacuum of
Eq.~(\ref{eq:susymin2}) for which we do not need to make strong
assumptions regarding the dimensionless coefficients $\omega_{i}$.  We
expect that there is a large region of parameter space (hence a large
number of UV completions) where the sEWSB vacua are physically
relevant.

%-----------------------------------------------------------------------------
\subsection{Supersymmetric Higgs Spectrum}
\label{sec:SUSYSpectrum}
%-----------------------------------------------------------------------------

The spectrum and interactions of the Higgs sector in the sEWSB vacuum
are particularly simple due to the constraints imposed by the unbroken
supersymmetry: the massive $W$ and $Z$ gauge bosons are components of
two separate massive vector superfields, a charged field with mass
$m_W$ and a neutral field with mass $m_Z$.  Each massive vector
superfield is made up of a massless vector superfield and an eaten
chiral superfield.  The \textit{complex} massive vector superfield
corresponding to the $W^\pm$ gauge bosons eats the superfields $H_u^+$
and $H_d^-$.  The massive vector superfield that contains the $Z$
boson eats the linear combination that does \textit{not} acquire a
VEV, $H \equiv (H_u^0 - H_d^0)/\sqrt{2}$.  The orthogonal combination
(or ``super-radial'' mode), $h \equiv (H_u^0 + H_d^0)/\sqrt{2}$,
remains as an additional degree of freedom and corresponds to the
physical Higgs superfield (the fact that $\langle h \rangle = v$
signals that these degrees of freedom are responsible for the
unitarization of $WW$ scattering).

The scalar components of the superfields, in unitary gauge, are
\bea
H_u = \pmatrix{H_u^+ \cr H_u^0} = 
\pmatrix{{1 \over \sqrt{2}}{H^+}\cr {v \over \sqrt{2}} + 
{1 \over 2}\left(H + h + i A^0\right)}~, 
\hspace{3mm} 
H_d = \pmatrix{ H_d^0 \cr H_d^-} = 
\pmatrix{{v \over \sqrt{2}} + {1 \over 2}\left(-H + h + i A^0\right)\cr {{1 \over \sqrt{2}}H^-}}~. 
\label{HiggsfluctuationsSUSY}
\eea
Here, $h$ is exactly the SM-like Higgs and we have decomposed the
scalar sector into mass eigenstates.  The scalar fields $H$ and
$H^{\pm}$ have masses $m_Z$ and $m_W$, respectively, and the fields
$h$ and $A^0$ ---in the zeroth order approximation discussed in the
previous subsection--- have mass $2 |\mu|$.\footnote{One can see that
the superfield $h$ has mass $2 |\mu|$ by using a supersymmetric gauge
transformation to completely remove the eaten {\it superfields} $H,
H^+_u, H^-_d$ from the theory.  The superpotential then contains the
mass term $W\supset \mu h^2$.} Also, the fermions of each eaten
superfield form Dirac partners with the vector superfield gauginos,
and have masses equal to their vector partners.  The Higgs
superpartner is a Majorana fermion.  The field content and
supermultiplet structure is as follows:

\begin{center}
\begin{tabular}{c|ccc}
Mass & Scalars & Fermions  & Vectors \\ \hline 
0 & --- & 1 majorana & $A_{\mu}$ \\
$m_W$ & $H^\pm$ & 2 Dirac & $W^{\pm}_{\mu}$\\
$m_Z$ & $H$ &  1 Dirac &  $Z_{\mu}$ \\
$2 |\mu|$ & $h,\ A^0$ & 1 majorana  & ---
\label{tab:components}
\end{tabular}
\end{center}

It is remarkable that in the sEWSB vacuum, the mass of the SM-like
Higgs (which completely unitarizes $WW$ scattering) is fixed by the
$\mu$-term.  In particular, the mass of the SM-like Higgs is
independent of the SM gauge couplings, contrary to what happens in the
MSSM with only renormalizable operators.  It should also be noted that
this mass can be shifted by order $\mu/\mu_{S}$ due to the tower of
higher-dimension operators.  The $H$ and $H^\pm$ masses remain tied to
the corresponding gauge boson masses, in the SUSY limit.

%-----------------------------------------------------------------------------
\subsection{Subleading Corrections, Canonical Normalization and Mixing}
\label{sec:mixing}  
%-----------------------------------------------------------------------------

As mentioned in Subsection~\ref{sec:EFT}, the \Kahler corrections
enter at second order in the $1/\mu_{S}$ expansion.  Such corrections
can affect both the spectrum and couplings of various fields, and
appear both through additional contributions to the scalar potential,
as well as through corrections to the kinetic terms.  It is
interesting that the former effects show up as a multiplicative factor
in the $F$-term potential.  As a concrete example, when the only
non-zero coefficient in the \Kahler potential of Eq.~(\ref{Kahler}) is
$c_{1}$, one finds the simple result
\beqa
V_{F} &=& \frac{|H|^{2}}{1+ \frac{c_{1}}{\mu^{2}_{S}} |H|^{2}} \,
\left|\mu + \frac{\omega_{1}}{\mu_{S}} H_{u} H_{d} + \cdots \right|^{2}~,
\label{NMMSSMPotential}
\eeqa
where $|H|^{2} \equiv H^{\dagger}_{u} H_{u} + H^{\dagger}_{d} H_{d}$.
This case arises precisely when the heavy physics corresponds to an
$SU(2)_{L} \times U(1)$ singlet (with $\kappa = 0$), as discussed in
Section~\ref{sec:UV}.  One can show that in the general case the first
factor is replaced by a real function $Z(H_{u},H_{d})$, whose exact
form is given in Eq.~(\ref{Zexact}) of
Appendix~\ref{app:kahlermetric}.  It follows that the \Kahler
corrections do not affect the vacuum obtained by imposing $F$-flatness
as if the \Kahler terms were of the minimal form.  However, they do
affect the spectrum and Higgs self-interactions, though such effects
are unlikely to be of immediate phenomenological relevance.

More relevant from a phenomenological point of view are certain
corrections to the Higgs kinetic terms, which are of order
$\mu/\mu_{S}$.  Although in the SUSY limit the properties of the
fields involved in the super-Higgs mechanism,
$\frac{1}{\sqrt{2}}(H^0_{u} - H^0_{d})$, $H^{+}_{u}$ and $H^{-}_{d}$,
are protected, those of the Higgs superfield itself can receive
important corrections.  For instance, the operator proportional to
$c_{1}$ in Eq.~(\ref{Kahler}) contains contributions to the kinetic
terms without the corresponding corrections to the gauge interactions
[in the sEWSB vacuum of Eq.~(\ref{eq:susymin2})]:
\beqa
\int \! d^{2}\theta d^{2}\bar{\theta} \, \frac{c_{1}}{\mu^{2}_{S}} \, |H_{u} H_{d}|^{2} &=& 
\frac{c_{1} v^{2}}{\mu^{2}_{S}} \left[ \frac{1}{2} \partial_{\mu} h \partial^{\mu} h +
\frac{1}{2} \partial_{\mu} A^0 \partial^{\mu} A^0 +
i \psi^{\alpha} \sigma^{\mu}_{\alpha\dot{\alpha}} \partial_{\mu} \bar{\psi}^{\dot{\alpha}}
\right] + \cdots~.
\label{kinetic mixing}
\eeqa
where we used the parametrization of Eq.~(\ref{HiggsfluctuationsSUSY})
and show only the kinetic terms, including those of the Higgs Majorana
partner.

The reason these effects are important is that, although formally of
second order in $1/\mu_{S}$, they correspond to the \textit{leading
order corrections} to the Higgs gauge interactions, after a rescaling
to restore canonical normalization:
\beqa
(h, A^0, \psi) \rightarrow \frac{1}{\sqrt{1+ \frac{2c_{1}\mu}{\omega_{1}\mu_{S}}}} (h, A^0, \psi) 
\approx \left(1- \frac{c_{1}\mu}{\omega_{1}\mu_{S}}\right) (h, A^0, \psi)~.
\label{rescalingSUSY}
\eeqa
Physically, these effects correspond
to mixing of the light fields with the UV physics at the scale
$\mu_{S}$.

%-----------------------------------------------------------------------------
\subsection{Non-renormalizable Operators at the Component Level}
\label{sec:HigherDPotential}  
%-----------------------------------------------------------------------------

So far we have emphasized the power-counting associated with operators
in the K\"ahler and superpotentials.  It is worth noting how the same
picture appears at the component level, especially since analyzing the
vacuum structure of the theory in the presence of SUSY breaking (as is
done in Section~\ref{sec:vacuumStructure}) requires a direct study of
the scalar \textit{potential}.

To zeroth order in $\mu/\mu_{S}$, and assuming for simplicity that
$\mu$ and $\omega_{1}$ are real, one gets an $F$-term potential with a
quartic interaction, as well as a certain ``dimension-6'' operator:
\beqa
V^{(0)}_{F} = \mu^{2} |H|^{2} + \frac{\omega_{1}\mu}{\mu_{S}} |H|^{2} (H_{u} H_{d} + {\rm h.c.}) + \frac{\omega^{2}_{1}}{\mu^{2}_{S}} |H|^{2} |H_{u} H_{d}|^{2}~,
\label{Potential0}
\eeqa
where $|H|^{2}$ was defined after Eq.~(\ref{NMMSSMPotential}).  The
quartic terms correspond to the $\lambda_{6}$ and $\lambda_{7}$
operators of the two-Higgs doublet model parametrization of
Refs.~\cite{Gunion:2002zf, Carena:2002es}.  The relevance of the
non-renormalizable term in Eq.~(\ref{Potential0}) depends on the
particular vacuum one is studying.  One should expand fields in
fluctuations around the relevant vacuum to determine which
interactions are important.  Since the sEWSB vacuum scales like
$\mu^{1/2}_{S}$, the ``dimension-6'' term should not be neglected: it
can contribute at the same order as the first two terms in
Eq.~(\ref{Potential0}).\footnote{In fact, it plays an essential role
in bounding the potential from below and stabilizing the vacuum of
interest; it also induces contributions to the quartic interactions of
the physical fluctuations about the sEWSB vacuum.} Thus, although it
should be obvious, we stress that the physics we are describing
\textit{cannot} be captured by the standard $SU(2)_{L}\times U(1)_{Y}$
two-Higgs doublet model parametrization based on renormalizable
interactions \cite{Brignole:2003cm}.

Similar comments apply at higher orders.  For instance, at first order
in the $\mu/\mu_{S}$ expansion, the operator proportional to $c_{1}$
in Eq.~(\ref{Kahler}) leads to additional quartic operators
(corresponding to $\lambda_{1}$, $\lambda_{2}$ and $\lambda_{3}$ in
the two-Higgs doublet model parametrization of Refs.
\cite{Gunion:2002zf,Carena:2002es}), to an additional ``dimension-6''
operator, and to a particular ``dimension-8'' operator, as can be
derived from Eq.~(\ref{NMMSSMPotential}):~\footnote{Note that, for
$c_{1} > 0$, $V^{(1)}_{F}$ can be large and negative, which would seem
to lead to a potential unbounded from below.  However, this occurs at
large values of the Higgs fields, where the EFT is not expected to be
valid.  Indeed, the remaining terms in the expansion of
Eq.~(\ref{NMMSSMPotential}) make the potential positive, as required
by SUSY.}
\beqa
V^{(1)}_{F} = -\frac{c_{1}\mu^{2}}{\mu^{2}_{S}} |H|^{4} - 
\frac{c_{1}\omega_{1}\mu}{\mu^{3}_{S}} |H|^{4} (H_{u} H_{d} + {\rm h.c.}) - 
\frac{c_{1}\omega^{2}_{1}}{\mu^{4}_{S}} |H|^{4} |H_{u} H_{d}|^{2}~.
\label{Potential1}
\eeqa
In spite of the different powers of $\mu_{S}$ in the denominators, all
of these can contribute to physical observables at first order in the
$\mu/\mu_{S}$ expansion in the sEWSB vacuum of
Eq.~(\ref{eq:susymin2}).  Nevertheless, our argument of
Subsection~\ref{sec:EFT}, performed at the level of the \Kahler and
superpotential, guarantees that the EFT around the sEWSB vacuum has a
well-defined expansion parameter and that the infinite tower of
operators can be consistently truncated, in spite of the
$\mu_{S}^{1/2}$ scaling of the sEWSB VEV.

In the next section we consider the effects of SUSY breaking at
tree-level.  However, we notice here that although loop effects from
supersymmetric partners can --in the presence of SUSY breaking-- give
contributions to the operators that play a crucial role in the
determination of the sEWSB vacuum, these are expected to be
subdominant.  For instance, the one-loop contributions to the
$\lambda_{6}$ and $\lambda_{7}$ quartic couplings are not
logarithmically enhanced and are proportional to
$A_{t}$~\cite{Haber:1993an}.  If all SUSY breaking parameters are of
order the EW scale, the corresponding one-loop contribution are of
order $3y^4_{t}/(16\pi^2)$ or smaller, which can easily be subdominant
compared to the quartic coupling in Eq.~(\ref{Potential0}) for
$\mu_{S} \sim (5-10) \mu$, as we envision here.  We therefore do not
consider loop effects any further and restrict ourselves to a
tree-level analysis.

%-----------------------------------------------------------------------------
\section{Supersymmetry Breaking}
\label{sec:SUSYBreaking}  
%-----------------------------------------------------------------------------

The previous section focused on electroweak symmetry breaking
\textit{in the SUSY limit}.  Although this limit is not fully
realistic, it allows a simple understanding of several properties of
the physics when SUSY breaking is taken into account.  Here we
reconsider the analysis including SUSY breaking effects.  SUSY
breaking terms are required, among other reasons, to lift the mass of
the photino.  They also break the degeneracy between the origin and
the non-trivial EWSB minimum.

%-----------------------------------------------------------------------------
\subsection{Scalar Potential}
\label{sec:potential}  
%-----------------------------------------------------------------------------

Our main assumption is that the heavy threshold, $\mu_{S}$, is very
nearly supersymmetric, so that a spurion analysis is
appropriate.\footnote{However, SUSY breaking in the heavy physics
sector can be of the same order as in the MSSM Higgs sector.  These
soft masses, together with the $\mu$-term, are assumed to be
parametrically smaller than $\mu_{S}$, which ensures that the EFT
analysis holds.} To order $1/\mu_S$, we must include the effects of
the non-renormalizable operator
\bea
W &\supset& {1 \over 2 \mu_S} \tilde{X} (H_u H_d)^2~,
\label{SUSYBreakingOp}
\eea
in addition to the usual soft terms in the MSSM Lagrangian, where
$\tilde{X} = \theta^2 m_{\rm soft}$ parameterizes the effective soft
SUSY breaking effects coming from the heavy sector.  We write, for
convenience, $m_{\rm soft} = \xi \omega_{1} \mu$, and assume that
$|\xi \omega_{1}| \lsim {\cal O}(1)$.  Thus, the relevant SUSY
breaking terms in the scalar potential read
\bea 
V_{\rm SB} &=& m^2_{H_u} |H_u|^2 +
m^2_{H_d} |H_d|^2 + \left[b \, H_u H_d - 
\xi \left({ \omega_1 \mu \over 2  \mu_s}\right) (H_u H_d)^2 + h.c. \right]~,
\nn 
\label{SUSYBreaking} 
\eea
and the potential to lowest order in the $1/\mu_{S}$ expansion takes
the form
\beqa
V &=& V_{\rm SB} + V_{D} + |H|^{2} \, \left|\mu + 
\frac{\omega_{1}}{\mu_{S}} H_{u} H_{d} \right|^{2}~,
\label{OurPotential}
\eeqa
where $|H|^{2}$ was defined after Eq.~(\ref{NMMSSMPotential}).  The $D$-term
potential is as in the MSSM:
\beqa V_{D} &=& \frac{1}{8} \, (g^{2} + g^{'2}) \left( |H^0_{u}|^{2}
- |H^0_{d}|^{2} + |H^{+}_{u}| - |H^{-}_{d}|
\right)^{2} + \frac{1}{2} \, g^{2} \left| H^{+}_{u} H^{0\dagger}_{d} +
H^{-\dagger}_{d} H^0_{u} \right|^{2}~.
\label{MSSMDtermPotential}
\eeqa

We start by considering the minimization of the potential,
Eq.~(\ref{OurPotential}).  Using $SU(2)_L$ transformations, we can
take $\langle H_{u} \rangle = (0,v_{u})$, with $v_{u}$ real, without
loss of generality.  By redefining the phase of $H^0_{d}$ we can then
take, as in the previous section, $\mu \mu_S / \omega_1$ real and
positive.  Note that the phases of $b$ and $\xi \mu^{2}$ are then
physical observables.\footnote{In the MSSM without higher-dimension
operators, it is customary to use the field reparameterization freedom
to choose $b$ real and positive.  We find it more convenient, when
studying the new vacua, to choose $\mu \mu_{S}/\omega_{1}$ real and
positive.} For simplicity, we will assume in the following analysis
that these parameters are real.

We also concentrate in a region of parameter space where no
spontaneous CP violation occurs, which can be guaranteed provided
either
\beqa
\frac{b}{|\mu|^{2}} > 0
\hspace{1cm} 
\textrm{or}
\hspace{1cm}
\xi \mu^{2} > 0~.
\nonumber
\eeqa
The first condition ensures that all the solutions to the minimization
equations are real, while the second would ensure that any putative
complex solution is \textit{not} a minimum of the potential.  Although
the above are only sufficient conditions to avoid spontaneous CP
violation, they will be enough for our purpose.  The possibility of
spontaneous CP violation in the presence of the higher-dimension
operators, although quite interesting, is beyond the scope of this
work.  Furthermore, we also note that for real solutions to the
minimization equations there are no charge-breaking vacua, provided
only that $m^{2}_{H_{d}}$ is not too negative.  Further details are
given in Appendix~\ref{sec:chargeAndCP}.

From here on we restrict ourselves to regions of parameter space where
electromagnetism is unbroken and CP is preserved, so that $\langle
H^0_{u} \rangle = v_{u}$ and $\langle H^0_{d} \rangle = v_{d}$ are
always real.  Notice that, unlike in the MSSM without higher-dimension
operators, the sign of $\tan\beta = v_{u}/v_{d}$ is physical.
However, we still have a remaining $U(1)_Y$ gauge rotation that we use
to choose $v_d$ positive, though $v_u$ may be positive or negative.
These non-trivial extrema of the potential are described by $v^2 =
v^{2}_{u} + v^{2}_{d}$ and $-\pi/2 < \beta <\pi/2$, and must satisfy
\bea
s_{2 \beta} &=& { 2b - 4 |\mu|^2 \rho (\rho s_{2\beta} -1) \over m^2_{H_u} + m^2_{H_d} + 
2 |\mu|^2 (\rho s_{2 \beta}-1)^2 - 2 \xi \mu^2 \rho }~, 
\label{eq:extremas}\\
m_Z^2  &=& {m^2_{H_u} - m^2_{H_d} \over c_{2 \beta}} - \left[m^2_{H_u} +m^2_{H_d} + 
2 |\mu|^2 (\rho s_{2 \beta} -1)^2\right]  ~, 
\label{eq:extremaz} \\
v^2 &\equiv& \rho \left({2 \mu \mu_S \over  \omega_1} \right)~.
\label{vdef}
\eea
Here $m_Z^2$ should be considered a placeholder for $v^2$ according to
$m_Z^2 = (g^2 + g^{\prime 2})v^2/2$.  For given ultraviolet parameters
($m^2_{H_{u}}, m^2_{H_{d}}, b, \mu, \mu_s/ \omega_{1}, \xi$) there may
be more than one solution to the above equations where EWSB occurs, in
addition to the origin where EWSB does not occur.  With our
conventions, a valid solution must also have real and positive $\rho$.

\begin{figure}[t]
\centerline{ \hspace*{-1cm}
\includegraphics[width=0.487 \textwidth]{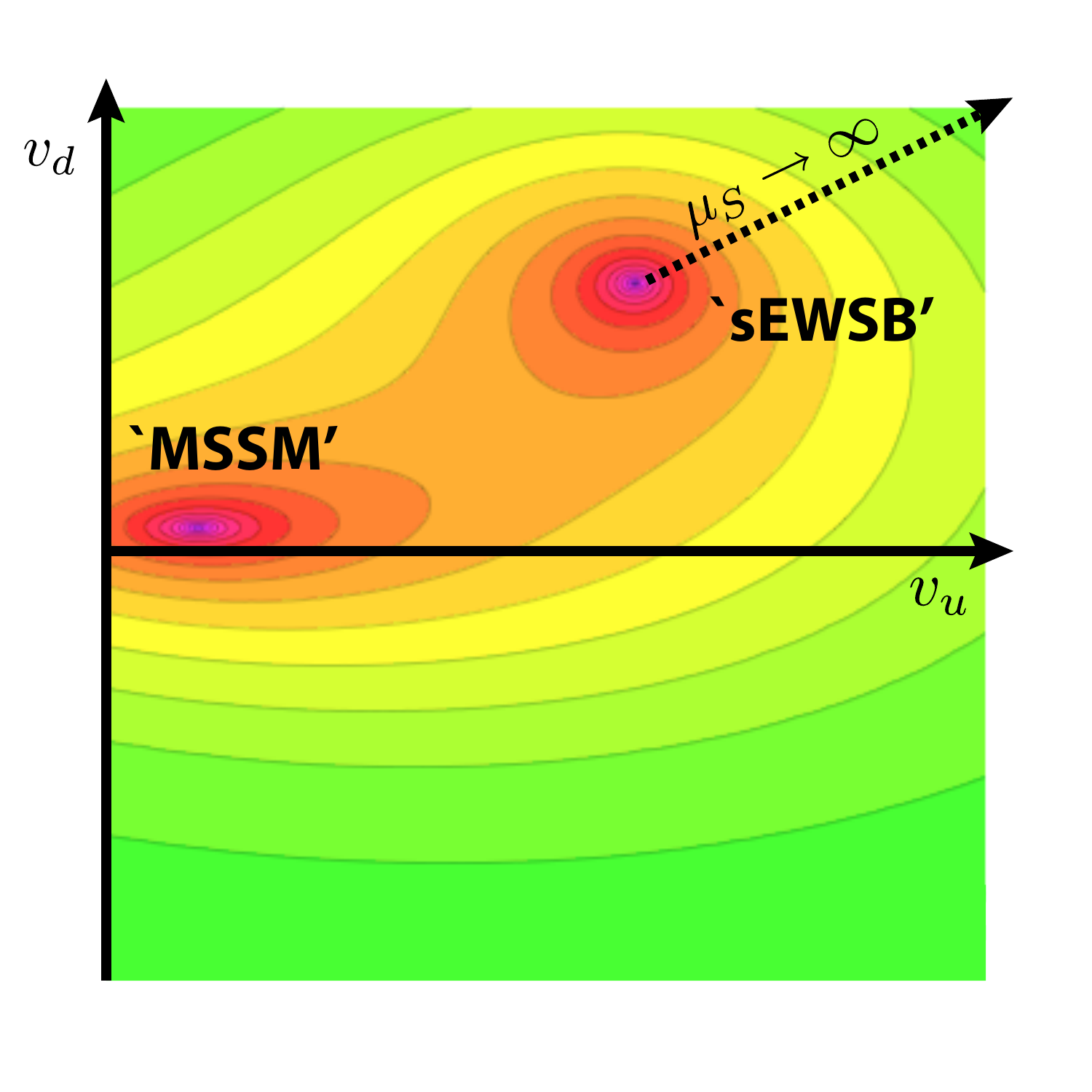}
} 
\caption{An illustration showing the equipotential lines in the
$v_{u}$--$v_{d}$ plane for a case with two nontrivial minima.  The
nature of these minima can be determined by exploring how the physics
depends on the UV scale $\mu_{S}$: the MSSM-like VEV remains near
the origin as $\mu_{S} \rightarrow \infty$, while the ``sEWSB'' VEV
scales like $\sqrt{\mu_{S}}$ (as indicated by the arrow) for large
$\mu_{S}$.  The limit is taken with all other microscopic parameters
fixed.}
\label{fig:twomin}
\end{figure}

The parameter $\rho$ introduced in Eq.~(\ref{vdef}) characterizes how
close these solutions are to the sEWSB minimum of
Section~\ref{sec:SUSYEWSB}: for vanishing soft parameters, one
recovers the SUSY expressions of the previous section, with $\rho
\rightarrow 1$ and $\tan \beta \rightarrow 1$.  On the other hand, the
MSSM-limit corresponds to $\rho \rightarrow 0$, or more precisely to
the scaling $\rho \rightarrow 1/\mu_{S}$ as $\mu_{S} \rightarrow
\infty$ [see Eq.~(\ref{vdef})].  This also suggests a definite
criterion to distinguish ---for finite $\mu_{S}$--- MSSM-like minima
from minima that involve the higher-dimension operators in a crucial
way.  While the VEV in an MSSM-like minimum tends to a constant as
$\mu_{S}$ becomes large, the new vacua are characterized by VEV's that
scale like $\sqrt{\mu_{S}}$ for large $\mu_{S}$, provided all other
microscopic parameters are kept fixed ($\rho$ remains of order one in
this limit).  This is illustrated in Fig.~\ref{fig:twomin}.
In other words, the new minima can be described as those that are
``brought in from infinity'' when the higher-dimension operators are
turned on.  It is important to notice that, as was argued by an
operator analysis in Subsection~\ref{sec:EFT}, the EFT gives a good
control of the physics of such non-standard vacua provided
\bea {v^2 \over \mu_S^2} \sim { 2 \rho \over \omega_1}{\mu \over \mu_S} \ll 1~.
\nn 
\eea
This approximation becomes even better in the limit described above
and leads to the interesting situation in which, although the physics
at $\mu_{S}$ is crucial in triggering EWSB, the \textit{details} of
that physics actually become unimportant.  With a slight abuse of
notation we will continue referring to vacua that obey the scaling $v
\sim \sqrt{\mu_{S}}$ in the large $\mu_{S}$ limit as sEWSB vacua, even
when SUSY breaking is not negligible.  The important property is that
they exist only due to the presence of the higher-dimension operators,
while being describable within the EFT framework.

%-----------------------------------------------------------------------------
\subsection{Higgs Spectrum}
\label{sec:higgsspectrum}
%-----------------------------------------------------------------------------

Besides studying the solutions to Eqs.~(\ref{eq:extremas}) and
(\ref{eq:extremaz}), which we will do in the next section, it is
important to determine their stability properties.  Here we work out
the Higgs spectrum in any extremum where electromagnetism is unbroken
and CP is conserved; the Higgs fields in the unitary gauge are
\bea H_u = \pmatrix{ c_{\beta}H^+ \cr v
s_{\beta} + {1 \over \sqrt{2}} \left(s_\alpha H^0+ c_\alpha h^0+ i c_{\beta} A^0 \right)\cr
}~, \hspace{5mm} H_d = \pmatrix{v c_{\beta} +  {1 \over \sqrt{2}}\left( c_\alpha H^0 - s_\alpha h^0 + i s_{\beta}A^0\right) \cr s_{\beta}H^-}~,
\label{Higgsfluctuations}
\eea
where $s_{\beta} = \sin\beta$, $c_{\beta} = \cos\beta$, etc.  For
arbitrary $(\rho,\beta)$, the charged and CP-odd Higgs masses are then
\beqa
m^{2}_{A^0} &=& {2 b \over s_{2 \beta}} + { 4 \rho |\mu|^{2} \over s_{2\beta}} + 
4 \rho \xi \mu^2 ~,
\label{m2A}
\\ [0.5em]
m^{2}_{H^{\pm}} &=& m^{2}_{W} + m^{2}_{A^0} - 4 \rho^{2} |\mu|^{2} - 2 \rho \xi \mu^2~,
\label{m2Hpm}
\eeqa
while the masses for the two CP-even scalars are given by
\beqa
m^{2}_{H^0,h^0} &=& \overline{m}^{2}  \pm \sqrt{\Delta m^{4} + m^{4}_{12}}~,
\label{m2hH}
\eeqa
where
\beqa
\overline{m}^{2} &=&  \frac{1}{2} m^{2}_{Z} + \frac{b}{s_{2\beta}} + 
\left( {2 \rho c_{4 \beta} \over s_{2 \beta}}+ 4 \rho^2 s_{2 \beta}^2 \right) |\mu|^2~,    
\nn \\
\Delta m^{2} &=& -\frac{1}{2} m^{2}_{Z} s_{2\beta} -b - 2\left( {3 \rho} - 
{4 \rho^2 s_{2 \beta}} \right) |\mu|^2 - 2 \rho \xi \mu^2 s_{2 \beta}~,
\label{hHmassmatrix} \\
m^{2}_{12} &=& - {\half m_Z^2 c_{2 \beta}} + b \cot_{2 \beta} + 
2 \rho |\mu|^2  \cot_{2 \beta}~.\nn
\eeqa
The mass mixing angle $\alpha$ satisfies 
\beqa
 \tan_{2 \alpha} =-\frac{\Delta m^{2}}{m^{2}_{12}}. 
\label{tanalpha}
\eeqa
The angle $\alpha$ is defined to agree with the two-Higgs doublet
model conventions for $m^2_{H^0} > m^2_{h^0}$ of
\cite{Gunion:2002zf,Carena:2002es}.  The SUSY-limit occurs as $\alpha
\rightarrow \pi/4$.  We note also that the $Z$-$Z$-$H^0$
($Z$-$Z$-$h^0$) coupling is proportional to $c_{\beta - \alpha}$
($s_{\beta - \alpha}$), where
\beqa
c^2_{\beta - \alpha} &=& {1 \over 2 (m_{H^0}^2 - m_{h^0}^2)} \left[ 3 (m_{H^0}^2 - m_{A^0}^2) + (m_{h^0}^2- m_Z^2) - 2 m_Z^2 s_{2\beta}^2 + 8 \rho \mu^2 s_{2 \beta} + 4 \rho \xi \mu^2 (c_{2\beta}^2+2) \right].
\nonumber
\eeqa
%

%-----------------------------------------------------------------------------
\subsection{Charginos and Neutralinos}
\label{sec:inos}
%-----------------------------------------------------------------------------

The chargino and neutralino spectra are also shifted from the SUSY
limit due to the presence of SUSY-breaking, in some cases (the
photino) drastically.  The shifts can be traced to multiple sources:
the presence of the Bino and Wino soft-masses ($M_1, M_2$), and the
shift of $(\rho, s_{2\beta})$ away from the SUSY-limit because of
soft-breaking in the Higgs scalar sector (see Subsection
\ref{sec:potential}).

The chargino mass matrix in the sEWSB vacuum is
\bea
{\cal L} &\supset& \left(\tilde{W}^+, \tilde{H}_u^+ \right) \left( \begin{array}{cc} M_2 & \sqrt{2} m_W c_{\beta} \\ \sqrt{2} m_W s_{\beta} & \mu \left( 1 - \rho s_{2 \beta} \right) \end{array} \right) \left(\begin{array}{c} \tilde{W}^- \\ \tilde{H}_d^- \end{array} \right).  \label{eq:charginomass}
\eea
In the SUSY-limit $(\rho, s_{2 \beta}) \rightarrow (1,0)$ and the pure
Higgsino entry in the chargino mass matrix vanishes; both charginos
become degenerate with the $W$ vector-boson.  In the more general case
with SUSY-breaking turned on, the eigenvalues are
\bea
m^2_{\chi_1,\chi_2} &=& {1 \over 2} M_0^2 \left\{ 1 \pm \sqrt{1 - {4 \left[ m_W^2 s_{2 \beta} - M_2 \mu \left( 1 - \rho s_{2\beta}\right) \right]^2 \over M_0^4}}\right\} \nn \\
M_0^2 &\equiv & \left[ M_2^2 + 2 m_W^2 + \mu^2 \left( 1 - \rho s_{2\beta}\right)^2 \right]. \nn 
\eea

The neutralino mass matrix in the sEWSB vacuum is
\bea
{\cal L} &\supset& \half \left(\tilde{B},\tilde{W}^3,   \tilde{H}_d^0, \tilde{H}_u^0\right) \left( \begin{array}{cccc} M_1 &  &- m_Z s_W c_{\beta} &m_Z s_W s_{\beta} \\ & M_2   & m_Z c_W c_{\beta} &-m_Z c_W s_{\beta}\\  -m_Z s_w c_{\beta}  &m_Z c_W c_{\beta}& 2 \mu \rho s_\beta^2 & -\mu \left( 1-2 \rho s_{2 \beta}\right) \\ m_Z s_W s_{\beta} &-m_Z c_W s_{\beta}&  -\mu \left( 1-2 \rho s_{2 \beta}\right) & 2 \mu \rho c_\beta^2 \end{array} \right)   \left(\begin{array}{c}\tilde{B}\\\tilde{W}^3\\   \tilde{H}_d^0\\ \tilde{H}_u^0 \end{array} \right), \nn \\
\label{eq:neutralinomass}
\eea
where $c_W$ stands for the weak-mixing angle $\cos \theta_W$.  A
massless neutralino with exactly the couplings of the photino emerges
from the spectrum in the SUSY-limit.

%-----------------------------------------------------------------------------
\subsection{Vacuum Structure}
\label{sec:vacuumStructure}
%-----------------------------------------------------------------------------

The presence of the higher-dimension operators in
Eqs.~(\ref{non-renorm-W}) and (\ref{SUSYBreakingOp}) lead to a rather
rich vacuum structure, even when restricted to the Higgs sector of the
theory.

\begin{figure}[t]
\centerline{ \hspace*{-1cm}
\includegraphics[width=0.487 \textwidth]{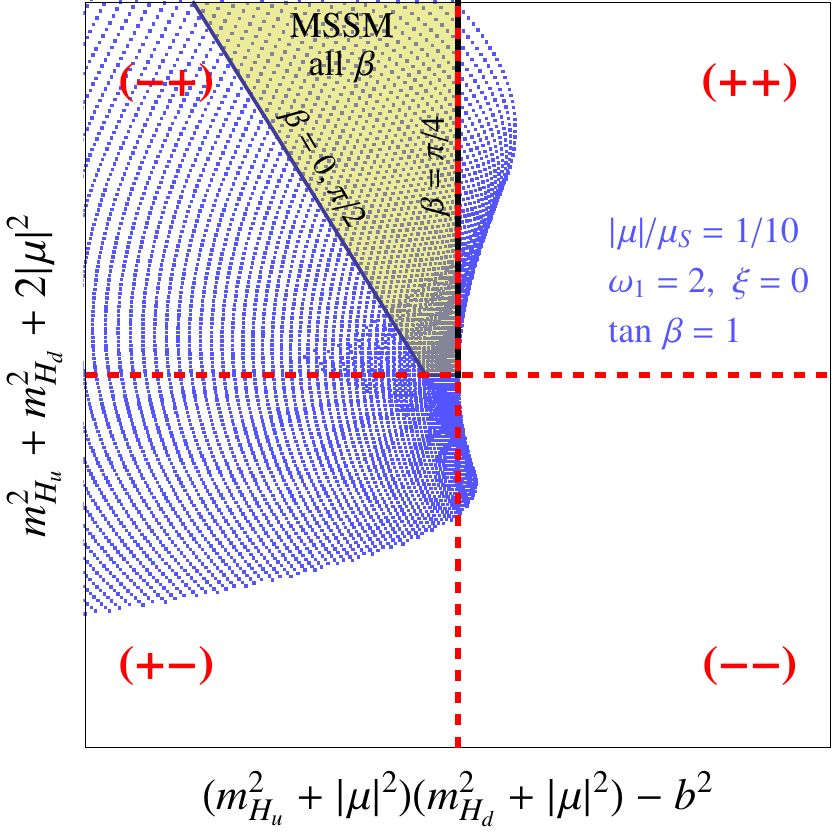}
} 
\caption{Region of parameters in the $({\rm det},{\rm trace})$ plane
of Eqs.~(\ref{dettrace}), that lead to EWSB. The light-shaded
triangular region corresponds to the complete EWSB parameter space in
the MSSM (in the absence of higher-dimension operators).  The (blue)
dots correspond to theories that break the EW symmetry, taking
$\omega_{1} = 2$, $\xi = 0$, and for fixed $\tan\beta = 1$
($m^2_{H_{u}} = m^2_{H_{d}}$).  We scanned over $b$ and $m^2_{H_{u}}$
with $|b|, |m^2_{H_{u}}| < (\mu_{S}/5)^2$.  All points have been
normalized so that $v = 174~{\rm GeV}$.}
\label{fig:phase}
\end{figure}

Let us start by recalling the situation in the MSSM without
higher-dimension operators.  The breaking of the EW symmetry can be
simply characterized by the behavior of the potential at the origin.
One considers the \textit{signs} of the determinant and trace of the
matrix of second derivatives (evaluated at the origin):
\beqa
{\rm det} &=& (m^{2}_{H_{u}}  + |\mu|^{2})(m^{2}_{H_{d}} + |\mu|^{2}) - b^{2}~,
\label{dettrace} \\
{\rm trace} &=& m^{2}_{H_{u}} + m^{2}_{H_{d}} + 2|\mu|^{2}~,
\nonumber
\eeqa
so that ${\rm sign} ({\rm det}, {\rm trace}) = (+,+)$ indicates that
the origin is a local minimum (the mass matrix squared has two
positive eigenvalues), while the other cases indicate that the origin
is unstable: $(+,-)$ is a maximum with two negative eigenvalues;
$(-,+)$ and $(-,-)$ indicate a saddle point with one negative and one
positive eigenvalue.  In the MSSM, the fact that all the quartic terms
arise from the $D$-terms, which have a flat direction along $|v_{u}| =
|v_{d}|$, leads to an additional constraint:
\beqa
m^{2}_{H_{u}} + m^{2}_{H_{d}} + 2|\mu|^{2} - 2 |b| > 0~,
\hspace{1cm}
\textrm{(MSSM stability)}
\label{stability}
\eeqa
which simply states that the \textit{quadratic} terms should be
positive along the flat direction.  This requirement eliminates the
cases $(-,-)$ and $(+,-)$ above [the trace is automatically positive,
hence it is not usually considered].  Using the MSSM minimization
conditions [Eqs.~(\ref{eq:extremas}) and (\ref{eq:extremaz}) with
$\rho = 0$], we can eliminate $b$ in favor of $\beta$ and $(m^2_{H_u}
- m^2_{H_d})$ in favor of $m^2_{Z}$, so that
\beqa
{\rm trace} = -\half m^2_{Z} - \frac{2\sec^22\beta}{m^2_{Z}} \, {\rm det}~, 
\hspace{1cm}
\textrm{(for MSSM)}
\eeqa
which shows that ``${\rm trace}$'' depends linearly on ``${\rm det}$''
with a $\beta$-dependent slope.  In addition, due to
Eq.~(\ref{stability}), for EWSB only the region ${\rm sign} ({\rm
det}, {\rm trace}) = (-,+)$ should be considered.  We show this
triangular region (light color) in Fig.~\ref{fig:phase}.

When the higher-dimension operators are included, the region of
parameter space in the $({\rm det},{\rm trace})$ plane that leads to
EWSB is considerably enlarged.  To illustrate this, we also show in
Fig.~\ref{fig:phase} the region that leads to a non-trivial minimum
for fixed $\tan\beta = 1$ [which from Eq.~(\ref{eq:extremaz})
corresponds to $m^2_{H_u} = m^2_{H_d}$].  For simplicity, we took
$\mu/\mu_{S} = 1/10$, $\omega_{1} = 2$, $\xi = 0$, and scanned over
the other parameters, requiring that $|b|, |m^2_{H_{u,d}}| \lsim
(\mu_{S}/5)^2$ to make sure that the EFT analysis is reliable
throughout.  We see that not only are the four quadrants $(+,+)$,
$(+,-)$, $(-,+)$ and $(-,-)$ accessible, but also that the stability
condition (\ref{stability}) is no longer necessary.

More interestingly, there are regions with multiple physically
inequivalent EWSB minima.  This should be clear from our discussion of
the supersymmetric limit in Section~\ref{sec:SUSYEWSB}, where we
pointed out that two degenerate minima exists (one that breaks the EW
symmetry and one that does not).  If a small amount of SUSY breaking
is turned on, such that the origin is destabilized, the minimum
initially at the origin can become non-trivial but remain near the
origin, while the originally sEWSB minimum is shifted only slightly.
The question then arises as to which of these two is the true global
minimum.  In the small SUSY breaking limit, this question is readily
answered by working out the shift in the potential energy to leading
order in the soft SUSY breaking terms:
\beqa
V \approx (m^2_{H_{u}} + m^2_{H_{d}} + 2b) \frac{v^{2}}{2}~,
\hspace{1cm}
\textrm{(small SUSY breaking)}
\label{globalmin}
\eeqa
where $v$ corresponds to the unperturbed SUSY VEV. For minima near the
origin, this result shows that its energy is not shifted at lowest
order in SUSY breaking.  Furthermore, we learn that the sEWSB minimum
with $v \approx (2\mu\mu_{S}/\omega_{1})^{1/2}$ is the global minimum
provided $m^2_{H_{u}} + m^2_{H_{d}} + 2b < 0$, at least when these
parameters are small compared to $\mu$.

In the general case, when SUSY breaking is not necessarily small
compared to $\mu$ (but still assuming it is small compared to
$\mu_{S}$ so that the EFT gives a reasonably good description of the
physics), we can approach the problem as follows: both
Eqs.~(\ref{eq:extremas}) and (\ref{eq:extremaz}) are only quadratic in
$\rho$, but fairly complicated in $\beta$.  We can solve
Eq.~(\ref{eq:extremas}) to characterize all extrema by two
branches:~\footnote{To simplify this expression, we assume that $\mu$
is real, though this is not necessary.  The general case is obtained by
making $\mu^{2} \rightarrow |\mu|^{2}$ and $\xi \rightarrow \xi
\mu^{2}/|\mu|^{2}$.
\label{realmu}}
\beqa
\rho_{\pm}\left(\beta\right) &=& {(1+ \frac{1}{2} \xi s_{2 \beta} + 
s_{2 \beta}^2 ) \over s_{2 \beta} (2+ s_{2 \beta}^2)} 
\left[ 1 \pm \sqrt{1 - {s_{2\beta}(2+ s^2_{2\beta}) 
\over (1+ \frac{1}{2} \xi s_{2 \beta} + s_{2 \beta}^2)^2}\left\{s_{2 \beta}
\left(1 + { m^2_{H_{u}} + m^2_{H_{d}} \over 2\mu^2}\right) - {b \over \mu^2}\right\}}\right]~. 
\nn \\
\label{eq:branches} 
\eeqa
The sEWSB vacua may be found in either the $\rho_+$ or the $\rho_-$
branch, while MSSM-type vacua are always in the $\rho_-$ branch
and are characterized by $\rho \sim 1/\mu_S$ as $\mu_S \rightarrow
\infty$.

Just as in the limit of small SUSY breaking effects, it is possible to
find potentials that contain multiple, inequivalent, sEWSB and
MSSM-type vacua with potential barriers in between. A complete
description of the phase space as a function of input parameters is
difficult to obtain, but it is straightforward to find examples of
EWSB minima that violate standard MSSM-assumptions. For example, the
origin can be unstable and outside of the MSSM-required
light-triangular region in Figure~\ref{fig:phase}, but a non-trivial
sEWSB vacuum is the stable, global minimum of the theory due to the
physics at $\mu_S$. More interestingly, there are potentials with a
local MSSM-type minimum that is unstable to decay to an sEWSB global
minimum, or vice-versa. These structures may have interesting
implications for cosmology and the cosmological phase transition to
the EWSB vacuum.

%-----------------------------------------------------------------------------
\section{sEWSB Vacua: Phenomenology}
\label{sec:pheno}
%-----------------------------------------------------------------------------

In this section we begin a preliminary analysis of the phenomenology
of the sEWSB vacua.  As defined in Section~\ref{sec:potential}, the
sEWSB vacua are distinguished from MSSM-like vacua due to their
behavior as $\mu_S \rightarrow \infty$, with all other microscopic
parameters fixed.  The sEWSB vacua exhibit a qualitative difference
from MSSM-like vacua in this limit: since the sEWSB vacua depend on
the scale $\mu_S$ to generate electroweak symmetry breaking, $v^2/\mu
\mu_S$ tends toward a constant as $\mu_S \rightarrow \infty$, even in
the presence of SUSY-breaking.

%-------------------------------------------------------------------
\subsection{Inverted CP-even Scalars}
%-------------------------------------------------------------------

Collider experiments have put tight constraints on the parameter space
of the MSSM. These constraints are mainly due to the LEP-II bound of
114 GeV on the neutral CP-even state which has SM-like couplings to
massive vector Z bosons.  It is much more natural for sEWSB vacua to
satisfy the 114 GeV bound on the SM-like Higgs state, since sEWSB
vacua naturally have an inverted scalar sector: the heavy CP-even
state is SM-Higgs-like, and is subject to the LEP-II bounds, while the
light CP-even state is not SM-like, couples more weakly to Z bosons,
and is more difficult to observe.

\begin{figure}[t]  
\centerline{\includegraphics[width=7.5 in]{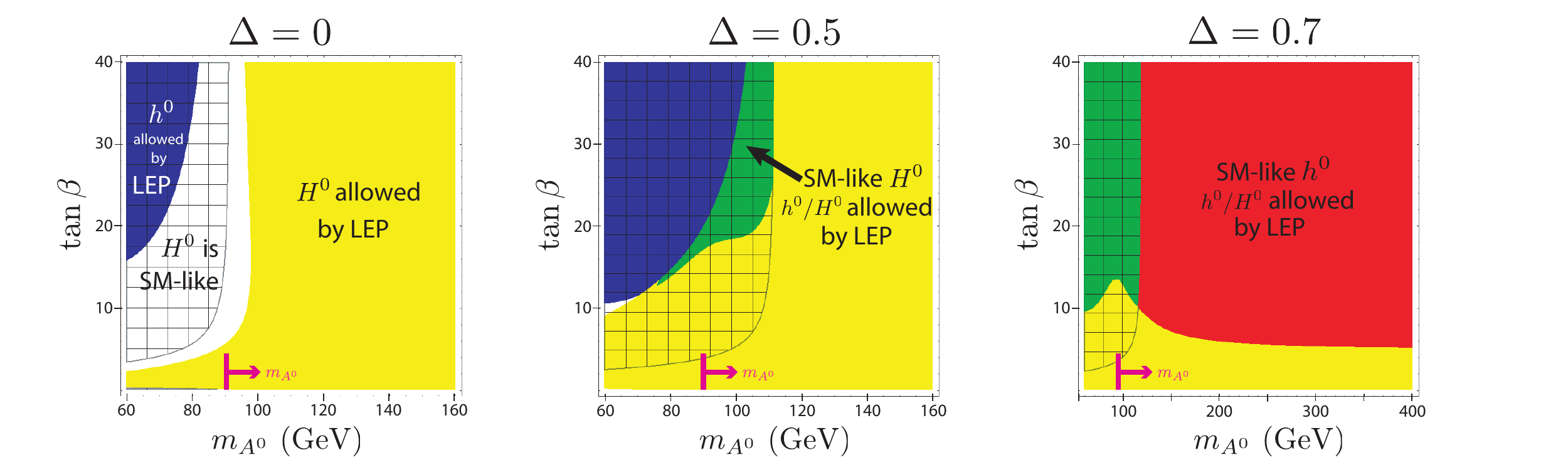}}
\caption{Inverted scalar hierarchy region in the MSSM, where the
heavier CP-even state $H^0$ is SM-like (hatched region), together with
the LEP-II allowed regions for $h^0/ H^0$ (blue/yellow)---with and
without quantum corrections from top-stop loops.  There is no viable
region with an inverted scalar hierarchy without quantum corrections
(leftmost plot).  Including a correction of size $\Delta=0.5$ (see
text) to the $H_u^0$--$H_u^0$ component of the CP-even neutral mass
matrix leads to a viable inverted scalar hierarchy (green region,
middle figure).  Setting $\Delta=0.7$ (right figure) produces both a
viable inverted scalar hierarchy region (green) and a viable standard
hierarchy region (red), where $h^0$ is SM-like.  These bounds include
quantum corrections only through their effects on the CP-even mixing
angle $\alpha$, and assume B$(h^0, H^0 \rightarrow b \bar{b}) \sim
0.85$.  The purple arrow indicates the LEP bound on $m_{A^0}$.}
\label{fig:inverted}  
\end{figure} 

Regions where the light CP-even state is not SM-like exist in the
MSSM, but are relatively rare and tuned \cite{Dermisek:2007ah}.  The
inverted hierarchy spectrum is distinct from the usual decoupling
limit of the MSSM, where an entire $SU(2)$ doublet of fields ($H^+,
H^0, A^0$) becomes much heavier than the weak-scale while the lighter
CP-even state $h^0$ is increasingly SM-like.  In
Figure~\ref{fig:inverted}, we qualitatively show in the
$m_{A^0}$--$\tan\beta$ plane the inverted hierarchy region (hatched)
where $H^0$ is more SM-like than $h^0$ (i.e. $g^2_{H^0
ZZ}/g^2_{h_{SM}ZZ} = c_{\beta - \alpha}^2 > 1/2$).  We use a smooth
interpolation of LEP-II bounds on the CP-even states only
\cite{Barate:2003sz} to describe regions of parameter space where
$h^0/H^0$ are allowed (blue/yellow regions).  We assume that all
superpartners are sufficiently heavy that no Higgs decay channels
other than the SM ones are open.  We take ${\rm B}(h^0, H^0
\rightarrow b \bar{b}) \sim 0.85$, which is the tree-level
approximation for $h^0$ and $H^0$ in the MSSM if the only important
decays are to tau and bottom pairs.\footnote{This assumption can hold
approximately beyond tree-level.  For instance, at large $\tan\beta$
these two decay channels are enhanced, and the branching fractions can
be close to the values used here even when quantum corrections are
included (see, for instance, Ref.~\cite{Carena:2002es}).  In the low
$\tan\beta$ region, decays of $H^{0}$ into $W$ pairs can be important,
but only when $m_{H^{0}}$ is above the $114~{\rm GeV}$ bound, so that
the LEP allowed regions are not expected to change.} The LEP
bound on $m_{A^0}$ of about $90~{\rm GeV}$~\cite{Abdallah:2003ip} is
indicated by the purple arrow in the plots. At tree-level
(leftmost panel in Figure~\ref{fig:inverted}) in the MSSM there is no
inverted hierarchy region that is compatible with LEP-II bounds.
Crucially in the inverted hierarchy region, $H^0$ has too large a
coupling to $Z$ bosons, while its mass is within 10\% of $m_Z$.

SUSY-breaking effects from top-stop loops create a narrow, viable
inverted hierarchy region (green region which is the overlap between
blue, yellow and hatched regions in the middle panel of
Figure~\ref{fig:inverted}).  We consider only quantum corrections from
the stop sector.  Inverted hierarchies occur in the MSSM at large
$\tan\beta$ whenever $m^2_{A^0} < m_Z^2 (1 + \Delta)$ (where $\Delta$
is the size of the quantum correction to the $H_u^0$--$H_u^0$
component of the neutral scalar mass matrix, normalized by $m_Z^2$
).\footnote{ For degenerate stops and small stop-mixing, the stop
masses must be close to 400 GeV to produce $\Delta \sim 0.5$, or 600
GeV to produce $\Delta \sim 0.7$.} As $\Delta$ increases, the hatched
region of Figure~\ref{fig:inverted} therefore begins to move to larger
$m_{A^0}$.  Meanwhile, $m^2_{H^0}$ grows in the inverted hierarchy
region ($\sim m_Z \sqrt{1 + \Delta}$) and begins to escape the LEP-II
bounds (its $Z$ couplings are relatively unaffected by $\Delta$).  The
lighter CP-even state is bounded from above by $m_{A^0}$, and the
effect of $\Delta$ is to reduce the couplings of $h^0$ to $Z$ bosons
(for fixed $m_{A^0}$ and $\tan\beta$ its mass is unaffected).
Therefore, the blue region where $h^0$ passes LEP constraints also
moves to heavier $m_{A^0}$.  This leads to a single region where both
experimental constraints overlap with the inverted scalar spectrum
(shown in green).  Although there is a viable inverted scalar
spectrum, $m^2_{A^0} \sim m^2_{H_d}- m^2_{H_u} - m_Z^2$ must be
satisfied to a high degree of accuracy in this region
\cite{Dermisek:2007ah}.

As is well known, if top-stop corrections are sufficiently large, a
region where $h^0$ is SM-like and escapes LEP-II bounds appears.  This
region is shown in red in the rightmost panel of
Figure~\ref{fig:inverted} for $\Delta=0.7$.  For sufficiently large
$\Delta$, this region is much larger than the viable inverted
hierarchy region where $H^0$ is SM-like. It is also possible that 
explicit CP-violation in the third generation squarks leads to a relaxation
of the LEP bounds on the MSSM Higgs sector at low and intermediate 
values of $\tan\beta$~\cite{Carena:2000ks}.

\begin{figure}[t]
\centerline{ %\hspace*{-1cm}
\includegraphics[width=1 \textwidth]{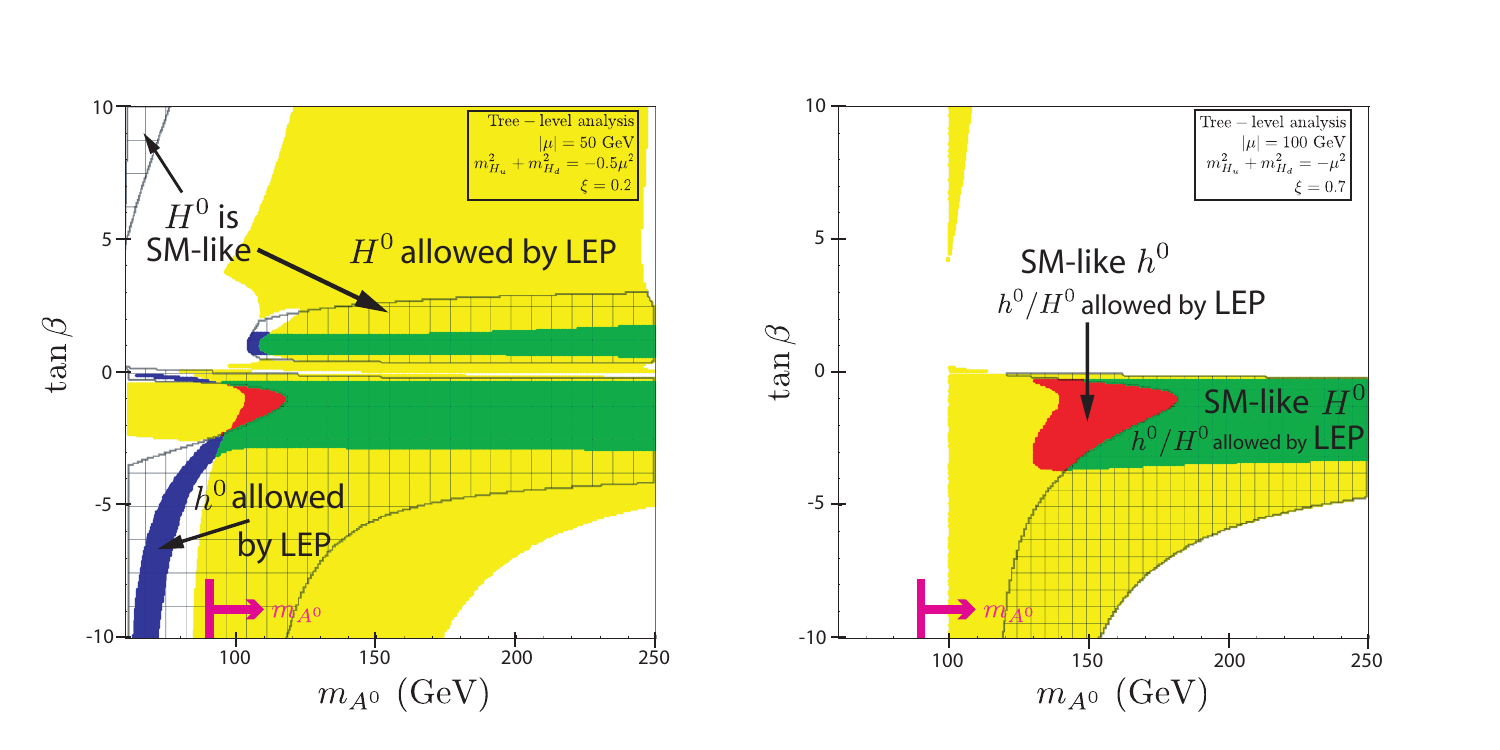}
} 
\caption{Examples illustrating the inverted hierarchy region in the
presence of non-renormalizable operators, as well as the regions
allowed by LEP. The color code is the same as in
Fig.~\ref{fig:inverted}.  The leading order tree-level expressions of
Section~\ref{sec:SUSYBreaking} are used, and no loop corrections are
included.  The charged Higgs direct bounds are satisfied in the LEP
allowed regions.  The purple arrow indicates the LEP bound on
$m_{A^0}$.  Direct limits on the lightest chargino/neutralino are not
shown.  The two plots correspond to different choices of the
parameters of the model other than $\tan\beta$ and $m_{A^0}$.}
\label{fig:ourmodel}
\end{figure}
In sEWSB vacua the scalar Higgs
properties can change significantly.  When the non-renormalizable operators of
Section~\ref{sec:SUSYBreaking} are included, the scalar Higgs
sector cannot be parameterized by $\tan\beta$ and $m_{A^0}$ alone, even
at tree-level.  As an illustration, we show in Fig.~\ref{fig:ourmodel}
two examples of the $m_{A^0}$--$\tan\beta$ plane that exhibit the
inverted CP-even scalar hierarchy region (hatched), fixing the values
of $|\mu|$, the sum $m^{2}_{H_{u}} + m^{2}_{H_{d}}$, and the SUSY
breaking parameter $\xi$ [the difference $m^{2}_{H_{u}} -
m^{2}_{H_{d}}$ is fixed by Eq.~(\ref{eq:extremaz})].  

We see that,
unlike in the MSSM, there exists a large, LEP allowed, inverted hierarchy region at
low $\tan\beta$. 
For reference, we also show the regions
allowed by the LEP Higgs searches in the CP-even sector, 
using the same color code as in Fig.~\ref{fig:inverted}.  We perform a tree-level analysis at leading order in  the $1/\mu_{S}$
expansion, ignoring loop corrections that depend on
additional SUSY breaking parameters (associated with the third
generation).  All the points we consider are within the domain
of validity of the EFT. We do not include in the plots the direct
chargino/neutralino exclusion limits, that are expected to impose
further constraints (see Section~\ref{charginos}); we have checked
that they do not change the qualitative picture shown in the plots.
These limits depend on the gaugino soft mass parameters that do not
enter in the scalar sector.  The neutralinos can be sufficiently heavy
for the bounds on the Higgs mass from invisible decays to be satisfied
in the regions marked as allowed in the plots.  We also assume that
the Higgs decays into $b\bar{b}$ are as important as in the MSSM (we
do not consider in this paper effects from physics beyond the MSSM
that affects particles other than those in the Higgs sector). The qualitative lesson is
that there are interesting new regions of parameter space that can be
consistent with existing limits, even at tree-level.  Furthermore,
this tends to happen for $|\tan\beta| = {\cal O}(1)$.

%-----------------------------------------------------------------------------
\subsection{sEWSB Vacua: The $|tan \beta| \sim 1$ Limit}
\label{smalltanbeta}
%-----------------------------------------------------------------------------

To better understand the features discussed in the previous
subsection, we take a $|\tan\beta| \sim 1$ limit, where the analytic
expressions in the scalar sector from Section~\ref{sec:higgsspectrum}
are more easily understood.  In the formulas of this section we
assume, for simplicity, that $\mu$ is real.\footnote{See Footnote
\ref{realmu} if the complex $\mu$ expression is needed.} Writing
$\tan\beta = \pm 1+ 2\delta\beta$, the extrema conditions of
Eqs.~(\ref{eq:extremas}) and (\ref{eq:extremaz}) reduce to
\bea
\rho_{\epsilon} &=& {\half {\xi} \pm 2 \over 3} \left\{ 1 + \epsilon \sqrt{1 - \frac{3}{(\half \xi \pm 2)^{2}}\left(1 + \frac{m^2_{H_{u}} + m^2_{H_{d}}}{2 \mu^2} \mp \frac{b}{\mu^{2}} \right) } \right\}~,  \nn \\
\delta\beta &=& \pm {m^2_{H_{d}} - m^2_{H_{u}} \over 2 \left(m_Z^2 + m^{2}_{H_{u}} + m^{2}_{H_{d}} + 2 \mu^2(1\mp \rho_\epsilon)^{2} \right)}~,
\nn
\eea
where the two branches discussed in
Subsection~\ref{sec:vacuumStructure} are labeled by $\epsilon = \pm$.

The neutral masses reduce to
\bea
m_{A^0}^2 &=& 4 (\pm1 + \xi) \rho \mu^2 \pm 2b~+ {\cal O}(\delta \beta^2), \nn \\
m^2_{H^0} &=& {1 \over 2} \left[ m_Z^2 + m_{A^0}^2+ 8 \mu^2 \rho (\rho \mp 1- \xi/2) + |D|\right]+ {\cal O}(\delta \beta^2), \nn \\
m^2_{h^0} &=& {1 \over 2} \left[ m_Z^2 + m_{A^0}^2+ 8 \mu^2 \rho (\rho \mp 1- \xi/2) - |D|\right]+ {\cal O}(\delta \beta^2), \nn \\
D &\equiv & m_Z^2 + m_{A^0}^2 - 8 \mu^2 \rho ( 2 \rho \mp 1 ).\nn
\eea
The mixing angle that determines whether $H^0$ ($c^2_{\beta - \alpha}
> 1/2$) or $h^0$ is SM-like ($c^2_{\beta - \alpha} < 1/2$) simplifies
considerably:
\bea
c^2_{\beta - \alpha} &=& \left\{ \begin{array}{cc} 0 + {\cal O}( \delta \beta^2) &
 D > 0   \\
1 + {\cal O}( \delta \beta^2 )&  D < 0   \end{array} \right..
\eea
It is easy to understand the result for the mixing angle $c^2_{\beta -
\alpha}$ (which is the coefficient of the $Z$-$Z$-$H^0$ coupling) for
$\tan\beta \sim 1$ by appealing to the SUSY limit of
Section~\ref{sec:SUSYEWSB}.  In the SUSY limit, the CP-even field with
mass $2 \mu$ is always the SM-like Higgs state.  When $D < 0$, it is
the heavy $H^0$ field whose mass reduces to the SUSY limit value of $2
|\mu|$ so $\cos^2_{\beta - \alpha} \rightarrow 1$.  When $D>0$, it is
the light $h^0$ field whose mass reduces to $2 |\mu|$ so
$\cos^2_{\beta- \alpha} \rightarrow 0$.

Finally, the charged Higgs mass $m^2_{H^+}$ is always very close to
the non-SM-like CP-even Higgs mass
\bea
m_{H^+}^2 &=& \left\{ \begin{array}{cc}  m^2_{H^0} + (m_W^2- m_Z^2)+ {\cal O}( \delta \beta^2 ) &
 D > 0   \\
m^2_{h^0} + (m_W^2- m_Z^2)+ {\cal O}( \delta \beta^2 ) &  D < 0   \end{array} \right..
\eea

In the $|\tan\beta| \sim 1$ limit with SUSY-breaking included, larger
$\rho$ always tends to push $D <0$ so that $H^0$ becomes the SM-like
Higgs state and we have an inverted hierarchy.  Up to corrections of
order $\delta \beta^2$, we see that the inverted hierarchy spectra is
consistent with LEP-bounds with only one condition: that the heavy
CP-even state $H^0$ has $m_{H^0} > 114$ GeV, and {\it no} condition on
the mass of the non-SM like CP-even state $h^0$.  Further, when the
inverted hierarchy holds, $m^2_{H^0} = 4\mu^2 \rho ( 3\rho \mp 2 -
\xi/2)$ which may easily be larger than $114$ GeV for moderate $\rho$
and $\mu$.  Recall from the previous subsection that one of the
reasons for the rarity of inverted hierarchies in the MSSM is the
difficulty of simultaneously satisfying LEP constraints on both
CP-even states.

The definition of sEWSB vacua given in Section~\ref{sec:potential}
allows us to see that sEWSB vacua typically have larger $\rho$, and
hence inverted spectra.  This is clear from the $\epsilon = +$ branch
in the expression for $\rho$, but it's also true in the $\epsilon=-$
branch.  Working in the EFT makes this clear: we require that $\mu^2_S
\gg \mu^2, m^2_{H_u}, m^2_{H_d}, b$ for the validity of the EFT. Given
these input parameters, the only trustworthy vacua where EWSB occurs
satisfy two generic relationships: $v^2 \sim \mu_S \mu$ or $v^2 \sim
\mu^2$ (with any other soft-mass possibly replacing $\mu$), depending
on whether the non-renormalizable terms proportional to $\mu_S$ help
stabilize the VEV or not.  The former case is exactly an sEWSB vacuum
by our criteria of Section~\ref{sec:potential}, and will have $\rho
\sim v^2/(\mu \mu_S) \sim {\cal O}(1)$, while the latter is an
MSSM-like vacua with $\rho \sim v^2/(\mu \mu_S) \sim (\mu/\mu_S)$.

As a complement to the qualitative picture exhibited in
Fig.~\ref{fig:ourmodel}, we give a couple of numerical examples (with
$|\tan\beta| \sim 1$) that illustrate the inverted hierarchy spectrum,
together with the charged Higgs and chargino/neutralino masses.  It
should be recalled that these numbers are expected to be accurate to
approximately ${\cal O}\left(v^2/\mu_S^2\right)$.  To be conservative,
we require that charginos are heavier than the kinematic reach at
LEP-II, $m_{\chi^+}> 104$ GeV, and that neutralinos are heavier than
half of the Z-mass: $m_{\chi^0} > 45$ GeV. Depending on the
composition of the charginos and neutralinos in terms of the
underlying Higgsino and gaugino states, these bounds may be
 relaxed \cite{Amsler:2008zz}.

We also require that the charged Higgses have mass greater than the
direct LEP-II search bound of $80$ GeV~\cite{Amsler:2008zz}.  There
are more stringent constraints from the Tevatron on charged Higgs
masses for low $\tan \beta$ when $m_{H^+} < m_t - m_b$.  For $\tan
\beta \sim 1$, $m_{H^+} \gsim 110$ GeV~\cite{Amsler:2008zz}.  These
searches ignore the possibility that the charged Higgs can decay to a
chargino/neutralino, which may alter the limits.  Additionally there
are strong indirect constraints, $m_{H^+} > 295$ GeV from the measured
rate of $b \rightarrow s \gamma$ \cite{Misiak:2006zs}, although
additional NNLO corrections appear to weaken this bound
\cite{Becher:2006pu}.  These indirect analyses assume no other sources
of new physics beyond the charged Higgs itself.  However, given that
the chargino tends to be light in this theory and is known to
interfere with the charged Higgs contribution to $b \rightarrow s
\gamma$~\cite{Barbieri:1993av}, and the spectrum of squarks (which may
also interfere with the charged Higgs contribution) is undetermined,
we restrict ourselves to considering only the direct charged Higgs
bound.

The following sample points have inverted scalar hierarchies, a wide
range of $m_{H^0}$, and different $Z$-$Z$-$H^0$ couplings:

{\bf Point 1}
\begin{center}
\begin{tabular}{|c|c|c|c|c|c|c|c|c|}\hline
$\mu$ & $\omega$ & $\mu/ \mu_s$ & ${b / \mu^2}$ & ${m^2_u / \mu^2}$ & ${m^2_{H_d} /  \mu^2}$ & $\xi$& $ M_1 / \mu$ & $M_2 / \mu$  \\ \hline
-60 & 1 & 0.11 & -2.2 & -1.7 & -0.60& 0.20 & 1.5 & 1.7 \\ \hline
\end{tabular}
\end{center}
\begin{center}
\begin{tabular}{|c|c|c|c|c|c|c|c|c|}\hline
$\rho$& $\tan{\beta}$ & $m_{h^0}$ & $m_{H^0}$ & $g^2_{H^0 ZZ}/ g^2_{h_{\rm SM}ZZ}$ & $m_{A^0}$ &$m_{H^+}$& $m_{\chi^+}$ & $m_{\chi^0}$ \\ \hline
0.47 & -1.3 & 120 & 150 & 0.98 &  100 & 120 & 110 & 90 \\ \hline
\end{tabular}
\end{center}
This is a spectrum where $H^0$ is SM-like, but its mass is well-above
the LEP-II limit, and well-above the mass of $h^0$.

{\bf Point 2}
\begin{center}
\begin{tabular}{|c|c|c|c|c|c|c|c|c|}\hline
$\mu$ & $\omega$ & $\mu/ \mu_s$ & ${b / \mu^2}$ & ${m^2_u / \mu^2}$ & ${m^2_{H_d} /  \mu^2}$ & $\xi$& $ M_1 / \mu$ & $M_2 / \mu$  \\ \hline
-150 & 2 & 0.14 & -1.1 & -0.99 & -0.51& 0.20 & 0.36 & 0.57 \\ \hline
\end{tabular}
\end{center}
\begin{center}
\begin{tabular}{|c|c|c|c|c|c|c|c|c|}\hline
$\rho$& $\tan{\beta}$ & $m_{h^0}$ & $m_{H^0}$ & $g^2_{H^0 ZZ}/ g^2_{h_{\rm SM}ZZ}$& $m_{A^0}$ &$m_{H^+}$& $m_{\chi^+}$ & $m_{\chi^0}$ \\ \hline
.20 & -1.3 & 190 & 210 & 0.77 &  185 & 190 & 105 & 60 \\ \hline
\end{tabular}
\end{center}
Point 2 is similar to point 1, but all the scalar masses (including
$m_{H^+}$) are closer to 200 GeV. $H^0$ is not entirely SM-like.

{\bf Point 3}
\begin{center}
\begin{tabular}{|c|c|c|c|c|c|c|c|c|}\hline
$\mu$ & $\omega$ & $\mu/ \mu_s$ & ${b / \mu^2}$ & ${m^2_u / \mu^2}$ & ${m^2_{H_d} /  \mu^2}$ & $\xi$& $ M_1 / \mu$ & $M_2 / \mu$  \\ \hline
-70 & 3.5 & 0.19 & 1.95 & -0.45 & -0.47& 0.70 & -1.0 & .86 \\ \hline
\end{tabular}
\end{center}
\begin{center}
\begin{tabular}{|c|c|c|c|c|c|c|c|c|}\hline
$\rho$& $\tan{\beta}$ & $m_{h^0}$ & $m_{H^0}$ & $g^2_{H^0 ZZ}/ g^2_{h_{\rm SM}ZZ}$ & $m_{A^0}$ &$m_{H^+}$& $m_{\chi^+}$ & $m_{\chi^0}$ \\ \hline
1.8 & 0.99 & 100 & 350 & 1 &  300 & 90 & 100 & 48 \\ \hline
\end{tabular}
\end{center}
Point 3 has a very heavy spectrum, due to the large value of $\omega$,
and --unlike Points 1 and 2-- it has $\tan\beta > 0$.  Note also that
$m_{h^0}$ and $m_{H^+}$ are nearly degenerate, and very split from
$m_{H^0}$ and $m_{A^0}$.

Values of $\tan\beta$ near one are not usually
considered in the MSSM, due to the LEP constraints on the CP-even Higgs states.  We see here that
this region is expected to be viable in a large class of
supersymmetric extensions.  For $|\tan\beta| \sim 1$
the top Yukawa coupling is $y_{t} \sim 1/\sin\beta \sim \sqrt{2}$, a
sizable enhancement compared to either the SM or the cases normally
considered in the MSSM. Since the couplings of the CP-even Higgses to
top pairs are $g_{ht\bar{t}}/g^{\rm SM}_{ht\bar{t}} \approx
\cos\alpha/\sin\beta$ and $g_{Ht\bar{t}}/g^{\rm SM}_{Ht\bar{t}}
\approx \sin\alpha/\sin\beta$ (assuming quantum corrections are not
particularly large), it is possible that the gluon-fusion Higgs
production cross section is enhanced compared to the SM.\footnote{Such
a large value of the top Yukawa coupling can lead to the loss of
perturbativity at high energies.  However, this would happen above the
new physics threshold at $\mu_{S}$, and it is a UV-dependent issue
that we do not address here (see further comments in
Section~\ref{sec:UV}).} Also, since a heavy SM-like CP-even scalar
$H^{0}$ can have a sizable branching fraction into $W$'s when its mass
is around the $WW$ threshold, the Tevatron may be starting to probe
the present scenario~\cite{TevatronHiggs}.

%-----------------------------------------------------------------------------
\subsection{Chargino NLSP}
\label{charginos}
%-----------------------------------------------------------------------------

In phenomenologically viable sEWSB vacua, it is important that the
lightest neutralino and lightest chargino have masses that are
significantly different from the SUSY-limit.  In the SUSY limit, the
lightest neutralino is the photino, which is massless, and the
lightest chargino is degenerate with the $W$ boson.  Adding the soft
mass $M_1$ raises the photino mass without much difficulty.  In the
SUSY-limit, the charged Higgsinos have no mass term, as can be seen
from the explicit expression for the chargino mass matrix in
Eq.~(\ref{eq:charginomass}).  Large $\mu (1-\rho s_{2 \beta})$ will
help lift the lightest chargino above the LEP-II bound.  This tends to
favor regions with negative $s_{2 \beta} < 0 $, and/or $\rho \neq 1$.

It may be the case that the effects of SUSY breaking lift the lightest
neutralino above the lightest chargino.  In a scenario with a
low-scale of SUSY-breaking, when the gravitino is the LSP, a chargino
NLSP may lead to a charged track that eventually decays into an
on-shell $W$ boson and missing-energy \cite{Kribs:2008hq}.  In the
example below, the chargino--neutralino mass difference is only on the
order of $5$--$10$ GeV which is approximately the size of additional
$\mu/\mu_S$ contributions from higher-order operators in the $1/\mu_S$
expansion that we have not considered.  The precise size of these
corrections can only be determined in a given UV completion.

{\bf NLSP Chargino}
\begin{center}
\begin{tabular}{|c|c|c|c|c|c|c|c|c|}\hline
$\mu$ & $\omega$ & $\mu/ \mu_s$ & ${b / \mu^2}$ & ${m^2_u / \mu^2}$ & ${m^2_{H_d} /  \mu^2}$ & $\xi$& $ M_1 / \mu$ & $M_2 / \mu$  \\ \hline
-70 & 1 & 0.11 & -1.6 & -1.7 & .22& 0.20 & 1.5 & 1.7 \\ \hline
\end{tabular}
\end{center}
\begin{center}
\begin{tabular}{|c|c|c|c|c|c|c|c|c|}\hline
$\rho$& $\tan{\beta}$ & $m_{h^0}$ & $m_{H^0}$ &$g^2_{H^0 ZZ}/ g^2_{h_{\rm SM}ZZ}$ & $m_{A^0}$ &$m_{H^+}$& $m_{\chi^+}$ & $m_{\chi^0}$ \\ \hline
0.34 & -1.8 & 120 & 140 & 0.82 &  110 & 125 & 100 & 110 \\ \hline
\end{tabular}
\end{center}

%-----------------------------------------------------------------------------
\section{Ultraviolet Scenarios}
\label{sec:UV}
%-----------------------------------------------------------------------------

So far we have restricted ourselves to an analysis of the low-energy
physics from an EFT point of view.  This has the advantage of making
more transparent (and also easier to analyze) the effects of the heavy
physics on the low-energy degrees of freedom (here the MSSM field
content) and allowed us to focus on the sEWSB vacua.

It is nevertheless worth pointing out that the tower of operators
involving only the MSSM Higgs superfields that we have considered [see
e.g. Eq.(\ref{non-renorm-W})] already arises in one of the simplest
extensions of the MSSM: the addition of a SM singlet.  To be more
precise, consider the renormalizable superpotential
\beqa
W = \mu H_{u} H_{d} + \lambda S H_{u} H_{d} + \frac{1}{2} \mu_{S} S^2 + \frac{\kappa}{3} S^3~.
\eeqa
If the singlet mass $\mu_{S}$ is sufficiently large, we can integrate
out $S$ using its supersymmetric equation of motion (we could keep the
SUSY covariant derivative terms)
\beqa
S = -\frac{1}{\mu_{S}} \left[ \lambda H_{u} H_{d} + \kappa S^2 \right]~.
\label{SEOM}
\eeqa
Replacing back in the superpotential and using the above equation of
motion iteratively, one gets the effective superpotential
\beqa
W_{\rm eff} = \mu H_{u} H_{d} - \frac{\lambda^2}{2\mu_{S}} (H_{u} H_{d})^2
-\frac{\lambda^3 \kappa}{3 \mu^3_{S}} (H_{u} H_{d})^3 + \cdots~.
\eeqa
The full tower of higher-dimension operators is generated with, in the
notation of Eq.~(\ref{non-renorm-W}), $\omega_{1} = -\lambda^2$,
$\omega_{2} = -\lambda^3 \kappa$, etc.  Note also that for $\kappa =
0$ only the lowest dimension operator, with coefficient $\omega_{1}$,
is generated.

Similarly, replacing Eq.~(\ref{SEOM}) in the minimal kinetic term for
the singlet, $S^\dagger S$, one generates the operator in
Eq.~(\ref{Kahler}) proportional to $c_{1}$, with $c_{1} =
|\lambda|^2$, as well as other higher-dimension operators whose
coefficients are proportional to $\kappa$.

The soft SUSY breaking operator considered in the EFT of the previous
sections can be generated from the following terms in the
superpotential:
\beqa
W &\supset& -\alpha_{1} X S H_{u} H_{d} - \frac{1}{2} \alpha_{2} \mu_{S} X S^{2}~, \nn 
\eeqa
where $\alpha_{1}$ and $\alpha_{2}$ are dimensionless coefficients,
and $X$ is a spurion that parameterizes SUSY breaking in the singlet
sector.  If these SUSY breaking effects are sufficiently small so that
the threshold at $\mu_{S}$ is approximately supersymmetric, we can
simply use Eq.~(\ref{SEOM}) to obtain the operator of
Eq.~(\ref{SUSYBreakingOp}), with the identification $\tilde{X} =
\lambda (2\alpha_{1} - \alpha_{2} \lambda) X$.

As illustrated in the sample points discussed in
Subsection~\ref{smalltanbeta}, we envision a case where
$\omega_{1}\sim 1 {\rm - few}$.  This is a result of the fact that the
weak scale in the sEWSB vacua arises as the geometric mean between
$\mu$ and $\mu_{S}$, that for phenomenological reasons $\mu$ cannot be
too small, and from the requirement that the EFT description be valid
[see Eqs.~(\ref{eq:susymin}) and (\ref{EFTvalidity})].  In the singlet
UV completion discussed in this section, we see that $\omega_{1}\sim 1 {\rm - few}$ corresponds
to $\lambda \sim 1-2$.  Thus, the fact that the lightest Higgs scalar
is heavier than in the MSSM can be understood as arising from a moderately large coupling.
In addition, the interesting new phenomenologically viable regions,
with $\tan\beta \sim 1$, also have a top Yukawa coupling $y_{t}$
slightly larger than one.  For $\lambda = y_{t} = \sqrt{2}$ and
$\kappa = 0$, the RG equations for the singlet theory above the scale
$\mu_{S}$ lead to a Landau pole around $100~{\rm TeV}$.  The presence
of such a Landau pole (as well as the issue of gauge coupling
unification) is a UV-dependent question.  Note, however, that we are
not required to assume strong coupling at the scale $\mu_{S}$.

Finally, we emphasize here that the EFT approach allows one to
consider more general scenarios than the addition of one singlet, even
if at the lowest order the singlet theory already induces all
operators considered in the detailed analysis of
Sections~\ref{sec:SUSYBreaking} and \ref{sec:pheno}.  The point is
that the next-to-leading order corrections can be different in other
UV completions that also generate the same lowest order operators.  In
general, the coefficients of operators of higher dimension need not
obey the correlations that follow from the identification between the
EFT and singlet theory coefficients discussed above.

%-----------------------------------------------------------------------------
\section{Conclusions}
\label{sec:outlook}  
%-----------------------------------------------------------------------------

Supersymmetric electroweak symmetry breaking (sEWSB) divorces LEP-II
constraints from the spectrum of CP-even masses in the most direct
route: the SM-like Higgs mass is not related to weak SM gauge
couplings.

We showed explicitly that sEWSB happens in the most general effective
theory describing the MSSM Higgs degrees of freedom.  We argued that
the sEWSB vacua can be consistently defined and captured within the
EFT, even in the presence of soft-terms that perturb the SUSY limit.
In particular, we showed that although higher-dimension operators play
a key role in the appearance of the sEWSB vacua, the physics is under
perturbative control and can be studied without the specification of a
UV completion.  This EFT captures any UV theory that has the following
properties: $i)$ a nearly supersymmetric threshold just above the weak
scale, $ii)$ physics beyond the MSSM that couples to the MSSM Higgs
superfields, and $iii)$ the MSSM low-energy field content.  The vacuum
structure of the theory is quite rich and may have interesting
cosmological consequences.

The EFT approach we use greatly simplifies the analysis of sEWSB
phenomenology.  We derived expressions for the low-energy spectrum
that generalize those of the MSSM with only renormalizable operators.
The sEWSB vacua naturally have an inverted scalar spectrum which is
more easily compatible with the LEP-II experimental constraints: it is
the heavier CP-even Higgs state that is SM-like, not the lighter.  We
also find that typically $\tan \beta \sim {\cal O}(1)$ in the sEWSB
vacua.  In the fermion sector, charginos may be lighter than
neutralinos, leading to NLSP chargino scenarios.  Further
phenomenological studies are needed to understand the full range of
collider signatures.

The most important open question deals with the coincidence of scales
in the theory.  Although the three important scales of the theory,
$\mu_S, \mu, m_S$ are separately technically natural, the clustering
of these scales suggests a common origin.  Only in the context of an
ultraviolet theory can one address whether there is a reason for
$\mu_S$ to be slightly above both the $\mu$ and soft-supersymmetry
breaking scales.

\vspace{5mm} 

{\bf Acknowledgements:} The authors thank M.~Carena, D.E.~Kaplan,
Y.~Nomura, T.M.P.~Tait and C.~Wagner, for insightful comments, and
especially T.M.P. Tait for carefully reading the manuscript.  PB
thanks the Aspen Center for Physics where part of this work was
completed.  PB and EP are supported by DOE grant DE-FG02-92ER40699.

\appendix

%-----------------------------------------------------------------------------
\section{Exact Scalar Potential  for an Arbitrary \Kahler Metric}
\label{app:kahlermetric}  
%-----------------------------------------------------------------------------

In this appendix, we consider the most general \Kahler potential,
without SUSY covariant derivatives, in a theory with two $SU(2)_{L}$
doublets, $H_{u}$ and $H_{d}$, with $U(1)_{Y}$ charges $-1/2$ and
$+1/2$, respectively.  This must be a real function of the
$SU(2)_{L}\times U(1)_{Y}$ invariants $H^{\dagger}_{u} \, e^{V}
H_{u}$, $H^{\dagger}_{d} \, e^{V} H_{d}$, $H_{u} \epsilon H_{d}$ and
$H^{\dagger}_{u} \epsilon H^{\dagger}_{d}$, where the $e^{V}$ factors
ensure gauge invariance, and $\epsilon$ is the $SU(2)_{L}$
antisymmetric two-index tensor, which we restore explicitly in this
appendix.  Notice that we employ a matrix notation and always write
$H_{u} \epsilon H_{d}$ and $H^{\dagger}_{u} \epsilon H^{\dagger}_{d}$
with $H_{u}$ and $H^{\dagger}_{u}$ to the left.  This makes it easier
to keep track of signs associated with these $SU(2)_{L}$ contractions.
In an expansion in a tower of operators suppressed by a large scale
$\mu_{S}$, the \Kahler potential takes the form given in
Eqs.~(\ref{Kahler}), (\ref{fu}) and (\ref{fd}).  The non-minimal
character of the \Kahler potential has to be taken into account when
deriving the scalar potential, which in the supersymmetric limit takes
the form $V = V_{D} + V_{F}$, where the first term arises from
integrating out the $D$-terms, while the second arises from the
$F$-terms.  Assuming that the gauge sector is described by the minimal
SUSY kinetic terms, $\int d^{2}\theta W^{\alpha}W_{\alpha} + {\rm
h.c.}$, the $D$-term potential takes the form
\beqa
V_{D} = \frac{1}{2} D_{2}^{a} D_{2}^{a} + \frac{1}{2} D_{1}^{2}~,
\label{VD}
\eeqa
where 
\beqa
D_{2}^{a} &=& \left. \frac{\partial K}{\partial (H^{\dagger}_{u} \, e^{V} H_{u})} \, H^{\dagger}_{u} \, \tau^{a} H_{u}
+ \frac{\partial K}{\partial (H^{\dagger}_{d} \, e^{V} H_{d})} \, H^{\dagger}_{d} \, \tau^{a} H_{d}  \right|_{V=0} 
\nonumber \\
D_{1} &=& \left. \frac{1}{2} \left( \frac{\partial K}{\partial (H^{\dagger}_{u} \, e^{V} H_{u})} \, H^{\dagger}_{u} H_{u}
- \frac{\partial K}{\partial (H^{\dagger}_{d} \, e^{V} H_{d})} \, H^{\dagger}_{d} H_{d} \right)  \right|_{V=0} 
\label{Ds}
\eeqa
with $\tau^{a}$ the $SU(2)_{L}$ generators, and
\beqa
\left. \frac{\partial K}{\partial (H^{\dagger}_{u} \, e^{V} H_{u})} \right|_{V=0} = 
1 + \frac{a^{u}_{1}}{\mu^{2}_{S}} H^{\dagger}_{u} H_{u} + \frac{a^{ud}_{1}}{\mu^{2}_{S}} H^{\dagger}_{d} H_{d} + \left( \frac{b^{u}_{1}}{\mu^{2}_{S}} H_{u} \epsilon H_{d} + {\rm h.c.} \right) + \cdots~,
\nonumber \\
\left. \frac{\partial K}{\partial (H^{\dagger}_{d} \, e^{V} H_{d})} \right|_{V=0} = 
1 + \frac{a^{d}_{1}}{\mu^{2}_{S}} H^{\dagger}_{d} H_{d} + \frac{a^{ud}_{1}}{\mu^{2}_{S}} H^{\dagger}_{u} H_{u} + \left( \frac{b^{d}_{1}}{\mu^{2}_{S}} H_{u} \epsilon H_{d} + {\rm h.c.} \right) + \cdots~.
\label{NonminimalDs}
\eeqa
In order to derive $V_{F}$ we need to invert the \Kahler metric, whose
components take the form
\beqa
g_{H^{\dagger}_{u}}^{\hspace{3mm}H_{u}} &\equiv& \partial_{H^{\dagger}_{u}} \partial_{H_{u}} K
\nonumber \\
&=& A_{0} + A_{1} H_{u} \, H^{\dagger}_{u} + A_{2} H_{u} \, (\epsilon H_{d}) + A_{3} (\epsilon H^{\dagger}_{d}) \, (\epsilon H_{d}) + A_{4} (\epsilon H^{\dagger}_{d}) \, H^{\dagger}_{u}~,
\nonumber \\
g_{H^{\dagger}_{u}}^{\hspace{3mm}H_{d}} &\equiv& \partial_{H^{\dagger}_{u}} \partial_{H_{d}} K
\nonumber \\
&=& B_{1} H_{u} \, (H_{u} \epsilon) + B_{2} H_{u} \, H^{\dagger}_{d} + B_{3} (\epsilon H^{\dagger}_{d}) \, H^{\dagger}_{d} + B_{4} (\epsilon H^{\dagger}_{d}) \, (H_{u} \epsilon)~,
\nonumber \\
g_{H^{\dagger}_{d}}^{\hspace{3mm}H_{u}} &\equiv& \partial_{H^{\dagger}_{d}} \partial_{H_{u}} K
\label{Kahlermetric} \\
&=& C_{1} H_{d} \, (\epsilon H_{d}) + C_{2} H_{d} \, H^{\dagger}_{u} + C_{3} (H^{\dagger}_{u} \epsilon) \, H^{\dagger}_{u} + C_{4} (H^{\dagger}_{u} \epsilon) \, (\epsilon H_{d})~,
\nonumber \\
g_{H^{\dagger}_{d}}^{\hspace{3mm}H_{d}} &\equiv& \partial_{H^{\dagger}_{d}} \partial_{H_{d}} K
\nonumber \\
&=& D_{0} + D_{1} H_{d} \, H^{\dagger}_{d} + D_{2} H_{d} \, (H_{u} \epsilon) + D_{3} (H^{\dagger}_{u} \epsilon) \, (H_{u} \epsilon) + D_{4} (H^{\dagger}_{u} \epsilon) \, H^{\dagger}_{d}~,
\nonumber
\eeqa
where the coefficients $A_{i}$, $B_{i}$, $C_{i}$ and $D_{i}$ are, in
general, field-dependent gauge invariant functions.  Notice also that
the hermiticity of the \Kahler metric implies that $A_{4} =
A^{*}_{2}$, $D_{4} = D^{*}_{2}$, $C_{1} = B^{*}_{3}$, $C_{2} =
B^{*}_{2}$, $C_{3} = B^{*}_{1}$ and $C_{4} = B^{*}_{4}$, while
$A_{0}$, $A_{1}$, $A_{3}$, $D_{0}$, $D_{1}$ and $D_{3}$ are real.  In
the above, we use a dyad notation such that, for example, $H_{u} \,
(\epsilon H_{d})$ is a $2\times 2$ matrix with components
$M^{\alpha}_{\hspace{2mm} \beta} = H^{\alpha}_{u} \, (\epsilon
H_{d})_{\beta} = H^{\alpha}_{u} \, \epsilon_{\beta \gamma}
H_{d}^{\gamma}$, where $\alpha$, $\beta$, $\gamma$ are $SU(2)_{L}$
indices.  The inverse metric, $\tilde{g}$, can be similarly expanded
in terms of gauge covariant quantities as
\beqa
\tilde{g}_{H^{\dagger}_{u}}^{\hspace{3mm}H_{u}} &=& \tilde{A}_{0} + \tilde{A}_{1} H_{u} \, H^{\dagger}_{u} + \tilde{A}_{2} H_{u} \, (\epsilon H_{d}) + \tilde{A}_{3} (\epsilon H^{\dagger}_{d}) \, (\epsilon H_{d}) + \tilde{A}_{4} (\epsilon H^{\dagger}_{d}) \, H^{\dagger}_{u}~,
\nonumber \\
\tilde{g}_{H^{\dagger}_{u}}^{\hspace{3mm}H_{d}} &=& \tilde{B}_{1} H_{u} \, (H_{u} \epsilon) + \tilde{B}_{2} H_{u} \, H^{\dagger}_{d} + \tilde{B}_{3} (\epsilon H^{\dagger}_{d}) \, H^{\dagger}_{d} + \tilde{B}_{4} (\epsilon H^{\dagger}_{d}) \, (H_{u} \epsilon)~,
\nonumber \\
\tilde{g}_{H^{\dagger}_{d}}^{\hspace{3mm}H_{u}} &=& \tilde{C}_{1} H_{d} \, (\epsilon H_{d}) + \tilde{C}_{2} H_{d} \, H^{\dagger}_{u} + \tilde{C}_{3} (H^{\dagger}_{u} \epsilon) \, H^{\dagger}_{u} + \tilde{C}_{4} (H^{\dagger}_{u} \epsilon) \, (\epsilon H_{d})~,
\label{invmetric}
\\
\tilde{g}_{H^{\dagger}_{d}}^{\hspace{3mm}H_{d}} &=& \tilde{D}_{0} + \tilde{D}_{1} H_{d} \, H^{\dagger}_{d} + \tilde{D}_{2} H_{d} \, (H_{u} \epsilon) + \tilde{D}_{3} (H^{\dagger}_{u} \epsilon) \, (H_{u} \epsilon) + \tilde{D}_{4} (H^{\dagger}_{u} \epsilon) \, H^{\dagger}_{d}~.
\nonumber
\eeqa
The coefficients $\tilde{A}_{i}$, $\tilde{B}_{i}$, $\tilde{C}_{i}$ and $\tilde{D}_{i}$ are found in a straightforward computation from 
\beqa
\sum_{j=u,d} \tilde{g}_{H^{\dagger}_{i}}^{\hspace{3mm}H_{j}} g_{H^{\dagger}_{j}}^{\hspace{3mm}H_{k}} &=& \delta_{ik}~.
\label{inversedefn}
\eeqa
The terms proportional to the identity give $\tilde{A}_{0} = 1/A_{0}$
and $\tilde{D}_{0} = 1/D_{0}$.  Further requiring that the
coefficients of the non-trivial $SU(2)_{L}$ invariants vanish, and
using $(H^{\dagger}_{u}\epsilon)^{\alpha} (H_{u}\epsilon)_{\alpha} =
H^{\dagger}_{u} H_{u}$ and $(\epsilon H^{\dagger}_{d})^{\alpha}
(\epsilon H_{d})_{\alpha} = H^{\dagger}_{d} H_{d}$, give four groups
of four equations each that can be solved for
$(\tilde{A}_{1},\tilde{A}_{2},\tilde{B}_{1},\tilde{B}_{2})$,
$(\tilde{A}_{3},\tilde{A}_{4},\tilde{B}_{3},\tilde{B}_{4})$,
$(\tilde{C}_{1},\tilde{C}_{2},\tilde{D}_{1},\tilde{D}_{2})$ and
$(\tilde{C}_{3},\tilde{C}_{4},\tilde{D}_{3},\tilde{D}_{4})$.

We record the solution when only the operators explicitly shown in
Eqs.~(\ref{Kahler})-(\ref{fd}) are included, assuming that all their
coefficients are real, and specializing, for simplicity, to the case
where $a^{u}_{1} = a^{d}_{1} = a^{ud}_{1} \equiv a_{1}$ and $b^{u}_{1}
= b^{d}_{1} \equiv b_{1}$:
\beqa
\tilde{A}_{1} &=& - \frac{1}{D} \left[ \frac{a_{1}}{\mu^{2}_{S}} + \frac{a_{1} c_{1} - b_{1}^{2}}{\mu^{4}_{S} A_{0}} \, |H|^{2} \right]~,
\nonumber \\ [0.5em]
\tilde{A}_{3} &=& - \frac{1}{D} \left[ \frac{c_{1}}{\mu^{2}_{S}} + \frac{a_{1} c_{1} - b_{1}^{2}}{\mu^{4}_{S} A_{0}} \, |H|^{2} \right]~,
\nonumber \\ [0.5em]
\tilde{A}_{4} &=& - \frac{1}{D} \left[ \frac{b_{1}}{\mu^{2}_{S}} - 2 \, \frac{a_{1} c_{1} - b_{1}^{2}}{\mu^{4}_{S} A_{0}} \, H_{u}\epsilon H_{d} \right]~,
\label{invmetriccoefficients}
 \\ [0.5em]
\tilde{D}_{0} &=& \tilde{A}_{0} = 1/A_{0}~,
\nonumber \\
\tilde{B}_{2} &=& \tilde{C}_{2} = \tilde{D}_{1} = \tilde{A}_{1}~,
\nonumber \\
\tilde{B}_{4} &=& \tilde{C}_{4} = \tilde{D}_{3} = \tilde{A}_{3}~,
\nonumber \\
\tilde{A}^{*}_{2} &=& \tilde{B}^{*}_{1} = \tilde{B}_{3} = \tilde{C}^{*}_{1} = \tilde{C}_{3} = \tilde{D}^{*}_{2} = \tilde{D}_{4} = \tilde{A}_{4}~,
\nonumber
\eeqa
where
\beqa
D &=& 3 A_{0}^{2} - A_{0} \left[ 2 + \frac{a_{1} - c_{1}}{\mu^{2}_{S}} \, |H|^{2} \right] 
+ \frac{a_{1} c_{1} - b_{1}^{2}}{\mu^{4}_{S}} \, \left( |H|^{4} - 4 |H_{u}\epsilon H_{d}|^{2} \right)~,
\nonumber \\ [0.5em]
&=& 1 + \frac{4a_{1} + c_{1}}{\mu^{2}_{S}} \, |H|^{2}
+ 5\, \frac{b_{1}}{\mu^{2}_{S}} \, (H_{u}\epsilon H_{d} + H^{\dagger}_{u}\epsilon H^{\dagger}_{d} ) + {\cal O}(H^{4}/\mu^{4}_{S})~,
\label{denominator}
\\ [0.5em]
A_{0} &=& 1 + \frac{a_{1}}{\mu^{2}_{S}} |H|^{2} + \frac{b_{1}}{\mu^{2}_{S}} (H_{u}\epsilon H_{d} + H^{\dagger}_{u}\epsilon H^{\dagger}_{d} )~,
\eeqa
and we used the short-hand notation $|H|^{2} = H^{\dagger}_{u} H_{u} +
H^{\dagger}_{d} H_{d}$.  The $F$-term potential can then be derived
from the superpotential, $W$, and inverse metric,
Eqs.~(\ref{invmetric}) and (\ref{inversedefn}), according to
\beqa
V_{F} &=& \sum_{i,j=u,d} \frac{\partial W}{\partial H_{i}} \, \tilde{g}_{H^{\dagger}_{i}}^{\hspace{3mm}H_{j}} \, \frac{\partial W^{\dagger}}{\partial H^{\dagger}_{j}}~.
\label{VF}
\eeqa
In general, the fields in the above potential are not canonically
normalized as a result of the non-minimal \Kahler terms, and this
should be taken into account when reading off physical properties such
as the spectrum.  However, the minima of the potential are not
affected by this.

In the same spirit as above, the superpotential can be expanded as a
power series in the holomorphic gauge invariant $H_{u}\epsilon H_{d}$
as
\beqa
W = \mu H_{u}\epsilon H_{d} + \sum^{\infty}_{n=1} \frac{1}{n+1} \frac{\omega_{n}}{\mu^{2n-1}_{S}} (H_{u}\epsilon H_{d})^{n+1}~,
\eeqa
which leads to the $F$-term potential (still with non-canonically
normalized kinetic terms)
\beqa
V_{F} &=& Z(H_{u},H_{d})
\left|\mu + \sum^{\infty}_{n=1} \frac{\omega_{n}}{\mu^{2n-1}_{S}} (H_{u}\epsilon H_{d})^{n} \right|^{2}~,
\eeqa
where $Z(H_{u},H_{d})$ is the real function
\beqa
Z(H_{u},H_{d}) &=&  |H_{d}|^{2} \left\{ \tilde{A}_{0} + \tilde{A}_{3} |H_{d}|^{2} + \left[ (\tilde{A}_{2} + \tilde{C}_{1}) H_{u}\epsilon H_{d} + {\rm h.c.} \right] \right\} 
\nonumber \\
& & \mbox{} + |H_{u}|^{2} \left\{ \tilde{D}_{0} + \tilde{D}_{3} |H_{u}|^{2} + \left[ (\tilde{B}_{1} + \tilde{D}_{2}) H_{u}\epsilon H_{d} + {\rm h.c.} \right] \right\} 
\nonumber \\
& & 
 \mbox{} + 2 ({\rm Re}\tilde{B}_{4}) |H_{u}|^{2} |H_{d}|^{2} + \left[ \tilde{A}_{1} + \tilde{D}_{1} + 2 \, {\rm Re}\tilde{B}_{2}\right] |H_{u}\epsilon H_{d}|^{2}~,
\label{Zexact}
\eeqa
and we used the relations among the $\tilde{A}_{i}$, $\tilde{B}_{i}$,
$\tilde{C}_{i}$, $\tilde{D}_{i}$ that follow from the hermiticity of
the inverse metric $\tilde{g}$ [see comment after
Eq.~(\ref{Kahlermetric})].  In the special case considered in
Eq.~(\ref{invmetriccoefficients}), we have
\beqa
Z(H_{u},H_{d}) &=& \frac{1}{D} \left\{ |H|^{2} \left[ 1 + \frac{b_{1}}{\mu^{2}_{S} } (H_{u}\epsilon H_{d} + H^{\dagger}_{u}\epsilon H^{\dagger}_{d} ) \right]  + 2 \, \frac{a_{1}}{\mu^{2}_{S} } (|H|^{4} - 2 |H_{u}\epsilon H_{d}|^{2} ) \right\} ~,
\eeqa
where $D$ is given in Eq.~(\ref{denominator}).  Setting $a_{1} = b_{1}
= 0$, leads to Eq.~(\ref{NMMSSMPotential}) in the main text.

It is also straightforward to include SUSY breaking effects that can
be parameterized by a spurion chiral superfield $X = \theta^{2} F_{X}$.
The contributions to the scalar potential can be written in terms of
the inverse \Kahler metric derived above.  Consider a \Kahler
potential of the form
\beqa
K(H_{i},H^{\dagger}_{j}) + X^{\dagger} K_{1}(H_{i},H^{\dagger}_{j}) + X K^{\dagger}_{1}(H_{i},H^{\dagger}_{j}) + X^{\dagger} X K_{2}(H_{i},H^{\dagger}_{j})~,
\label{SUSYBreakingKahler}
\eeqa
where $K$, $K_{1}$ and $K_{2}$ are arbitrary functions (except $K$ and
$K_{2}$ are real).  By using the $F$-term equations of motion, one
easily finds an $F$-term potential
\beqa
V_{F} = ( \partial_{H_{i}} W ) \tilde{g}_{H^{\dagger}_{i}}^{\hspace{3mm}H_{j}}( \partial_{H^{\dagger}_{j}} W^{\dagger})
+ \left[ F_{X} ( \partial_{H_{i}} W ) \tilde{g}_{H^{\dagger}_{i}}^{\hspace{3mm}H_{j}}( \partial_{H^{\dagger}_{j}} K^{\dagger}_{1}) + {\rm h.c.} \right]
+ F^{\dagger}_{X} F_{X} ( \partial_{H_{i}} K_{1} ) \tilde{g}_{H^{\dagger}_{i}}^{\hspace{3mm}H_{j}}( \partial_{H^{\dagger}_{j}} K^{\dagger}_{1})
\nonumber
\eeqa
that generalizes Eq.~(\ref{VF}) [sums over $i,j = u,d$ are implicit].
The inverse metric $\tilde{g}$ is given in Eq.~(\ref{invmetric}).  The
contribution to the potential from the last term in
Eq.~(\ref{SUSYBreakingKahler}) is simply $F_{X}^{\dagger} F_{X}
K_{2}(H_{i},H^{\dagger}_{j})$ with the fields $H_{i}$ interpreted as
the scalar components.  There are no new contributions to the $D$-term
potential.

%-------------------------------------------------------------------------------------------
\section{CP Violation and Charge-Breaking Minima}
\label{sec:chargeAndCP}
%-------------------------------------------------------------------------------------------

Consider the potential of Eq.~(\ref{OurPotential}) and look for minima
of the form $\langle H_{u} \rangle = (0,v_{u})$, $\langle H_{d}
\rangle = (v_{CB},v_{d} e^{i\delta})$, where $v_{u}$, $v_{d}$ and
$v_{CB}$ are real.  We can choose this form for $\langle H_{u}
\rangle$ by performing an appropriate $SU(2)_{L}$ rotation.  It is
also clear from the form of the potential that, having set $H^{+}_{u}
= 0$, it depends only on $|H^{-}_{d}| \equiv v_{CB}$.  Furthermore, as
discussed in the main text, we can assume that $\mu\mu_{S}/\omega_{1}$
is real and positive, while the phases of $b$ and $\xi\mu^{2}$ are
physically observable.  However, we will assume, for simplicity, that
these two phases vanish and establish simple conditions such that
spontaneous CP violation or charge-breaking minima do not occur.

The $\delta$-dependent part of the potential takes the form
\beqa
V \supset x \cos\delta +y \cos^{2}\!\delta~,
\eeqa
with
\beqa
x = -v^{2} s_{2\beta} \left[ b + 2 \rho |\mu|^{2}\, \frac{v^{2} + v^{2}_{CB}}{v^{2}} \right]~,
\hspace{1cm}
y = -v^{2} s^{2}_{2\beta} \,\rho \,\xi \mu^{2}~,
\eeqa
where $\rho > 0$ was defined in Eq.~(\ref{vdef}).  Hence, the
derivative w.r.t. $\delta$ vanishes either for $\sin\delta = 0$, or when
\beqa
\cos\delta = -\frac{x}{2y} = -\frac{|\mu|^{2}}{\xi \mu^{2} s_{2\beta}} 
\left[ \frac{v^{2} + v^{2}_{CB}}{v^{2}} + \frac{1}{2\rho} \frac{b}{|\mu|^{2}} \right]~.
\label{solCPviolation}
\eeqa
Since $|\cos\delta| \leq 1$, this solution is not always physical.  In
particular, it does not exist provided $b/|\mu|^{2} \geq 0$ and $\xi
\lsim {\cal O}(1)$ (for $\omega_{1} \sim {\cal O}(1)$, we are already
assuming this latter condition to ensure that the heavy physics
corresponds to an approximately supersymmetric threshold).  On the
other hand, the solution may be allowed if there is some degree of
cancellation between the two terms in the parenthesis.  In this case,
one should still check whether the extremum corresponds to a minimum
of the potential or not.  In particular, the second derivative with
respect to $\delta$, evaluated on Eq.~(\ref{solCPviolation}), is
\beqa
\frac{\partial^{2} V}{\partial \delta^{2}} = 2 y \left[ 1- \cos^{2}\delta \right]~,
\eeqa
which has the sign of $y$, hence the sign of $-\xi \mu^{2}$.
Therefore, if $\xi \mu^{2} > 0$ this solution cannot be a minimum, and
the minima must be described by real VEV's.  We always assume one of
these two simple, sufficient conditions, $b/|\mu|^{2} \geq 0$ or $\xi
\mu^{2} > 0$, in the main text.

With these conditions for real VEV's, we can address the issue of
dangerous charge-breaking minima, i.e. solutions with $v_{CB} \neq 0$.
Setting $\delta = 0$, and considering $\partial V/\partial v_{CB} = 0$
one can see that any solution with $v_{CB} \neq 0$ must satisfy
\beqa
v_{CB}^2 = - \frac{1}{(g^{2} + g^{\prime 2})}
\left\{ 4 m^2_{H_{d}} + v^2 \left( g^2 + g^{\prime2} c_{2\beta} \right) + 
4 |\mu|^{2}  \left(\rho s_{2\beta} - 1\right)^2 \right\}~.
\label{vCB2}
\eeqa
Except for $m^2_{H_{d}}$, all the terms in the braces are explicitly
positive (recall $g^\prime<g$).  Since $v^2_{CB}$ must be positive,
$m^2_{H_{d}} \geq 0$ (or not too negative) is a sufficient condition
to ensure that charge-breaking extrema do not exist.  However, we note
that even if (\ref{vCB2}) is positive, one must check that it is
compatible with the remaining extremization conditions, that any such
solution is indeed a minimum, and whether it is a global as opposed to
a local minimum.

%-------------------------------------------------------------------------------------------

\end{document}